\documentclass[aps,prb,twocolumn,superscriptaddress,floatfix,showkeys]{revtex4-2}
\usepackage{etoolbox}
\usepackage[english]{babel}
\usepackage{graphics,graphicx} 
\usepackage{natbib}
\usepackage[labelformat=parens]{subfig}
\usepackage{float}
\usepackage{placeins} 
\usepackage{ragged2e}  
\usepackage{caption}
\usepackage{xcolor}
\usepackage{amsmath}
\usepackage{amssymb}
\usepackage{braket}
\usepackage{blindtext}
\usepackage{bm}
\usepackage{algorithm}
\usepackage{algpseudocode}
\usepackage{hyperref}
\hypersetup{
    colorlinks=true,
    citecolor=blue, 
    linkcolor= red, 
    urlcolor= blue 
}
\usepackage[labelformat=parens]{subfig}
\usepackage[justification=raggedright, font=footnotesize]{caption}

\usepackage{MnSymbol}

\usepackage[cal=boondox,scr=boondoxo]{mathalfa}

\begin{document}

\title{Autoregressive Projective Quantum Monte Carlo: From a Hermitian to a Non-Hermitian Perspective}

\author{Lavoisier Wah}
\email[]{lavoisier.wahkenounouh@mpl.mpg.de}
\affiliation{Max Planck Institute for the Science of Light, 91058 Erlangen, Germany}
\affiliation{Department of Physics, Friedrich-Alexander-Universit\"at Erlangen-N\"urnberg, 91058 Erlangen, Germany}

\author{Remmy Zen}
\affiliation{Max Planck Institute for the Science of Light, 91058 Erlangen, Germany}
\affiliation{School of Physics and Astronomy, Monash University, Clayton VIC 3800, Australia}

\author{Flore K. Kunst}
\email[]{flore.kunst@mpl.mpg.de}
\affiliation{Max Planck Institute for the Science of Light, 91058 Erlangen, Germany}
\affiliation{Department of Physics, Friedrich-Alexander-Universit\"at Erlangen-N\"urnberg, 91058 Erlangen, Germany}

\date{\today}

\begin{abstract}
Accurately determining the ground-state properties of quantum many-body systems remains a central challenge. In this work, we introduce an autoregressive projective quantum Monte Carlo (PQMC) framework that leverages recurrent neural networks (RNNs) to guide the stochastic dynamics. By incorporating autoregressive sampling into PQMC, we demonstrate substantial improvements in accuracy compared to standard unguided PQMC, while retaining polynomial computational cost. We benchmark our approach against conventional variational RNN ans\"atze and find that the autoregressive PQMC consistently achieves lower energies and higher fidelity, regardless of system size or whether the Hamiltonian is Hermitian or non-Hermitian. Our results highlight the versatility and power of neural-guided PQMC methods, paving the way for promising scalable simulations of low-energy states in complex quantum many-body systems.
\end{abstract}

\maketitle

\section{Introduction}\label{S0}

Quantum Monte Carlo (QMC) encompasses a large family of computational methods, whose common aim is the study of complex quantum systems. One of the main goals of these approaches is to provide a reliable solution (or at least an accurate approximation) to the quantum many-body problem Schr\"odinger equation~\cite{anderson1975random, reynolds1982fixed, VONDERLINDEN199253,Britoquantum2019}. The various QMC approaches~\cite{ceperley1980ground,sprague2024variational,ceperley1977monte, herman1982path, inack2018phd, kramer2024quantum} all rely on the Monte Carlo (MC) method to handle the multidimensional integrals that arise in different formulations of the many-body problem. One of the most widely used QMC techniques is variational Monte Carlo (VMC)~\cite{ceperley1980ground,sprague2024variational}, which involves choosing a trial wavefunction with adjustable parameters and optimizing these parameters to minimize the energy expectation value. Another variant of QMC, which constitutes the raison d'\^etre of this work, is projective quantum Monte Carlo (PQMC) \cite{ceperley1995path,inack2015simulated,inack2018phd}, also often referred to as diffusion Monte Carlo (DMC)~\cite{carlo1990diffusion} or Green's function Monte Carlo (GFMC)~\cite{ceperley1995path}. This method goes beyond VMC by projecting the ground state from a trial wavefunction using a stochastic process that simulates the time evolution of the Schr\"odinger equation in imaginary time at zero temperature. PQMC is known as one of the most powerful tools for accurately simulating the ground state of many-body systems, outperforming VMC (which is upper bounded)~\cite{pilati2019self,inack2015simulated,inack2018phd,Feldbacher2004ProjectiveQM}.

In principle, any physical system can be described by the many-body Schr\"odinger equation as long as the constituent particles are not moving ``too fast''. Solving the Schr\"odinger equation for a given system enables the prediction of its physical behavior. The main difficulty, however, is that its solution requires knowledge of the many-body wavefunction in the many-body Hilbert space, which typically grows exponentially with the number of particles. As a consequence, solving the Schr\"odinger equation for a very large number of particles is generally impossible within a reasonable amount of time, even with modern parallel computing technology. Traditionally, approximations to the many-body wavefunction have been expressed as (anti)symmetric functions of one-body orbitals~\cite{parr1982density} in order to obtain a tractable formulation of the Schr\"odinger equation. However, such formulations suffer from several drawbacks, either limiting the effects of quantum many-body correlations, as in the case of the Hartree–Fock approximation, or exhibiting very slow convergence~\cite{isozaki2004many, march1995many}. QMC provides a direct way to study the many-body problem and the many-body wavefunction beyond these approximations~\cite{neklyudov2024wasserstein}. QMC is particularly well suited for systems in which traditional approaches, such as perturbation theory or mean-field approximations, fail due to strong correlations.

\begin{figure}
    \centering
    \includegraphics[width=0.5\textwidth]{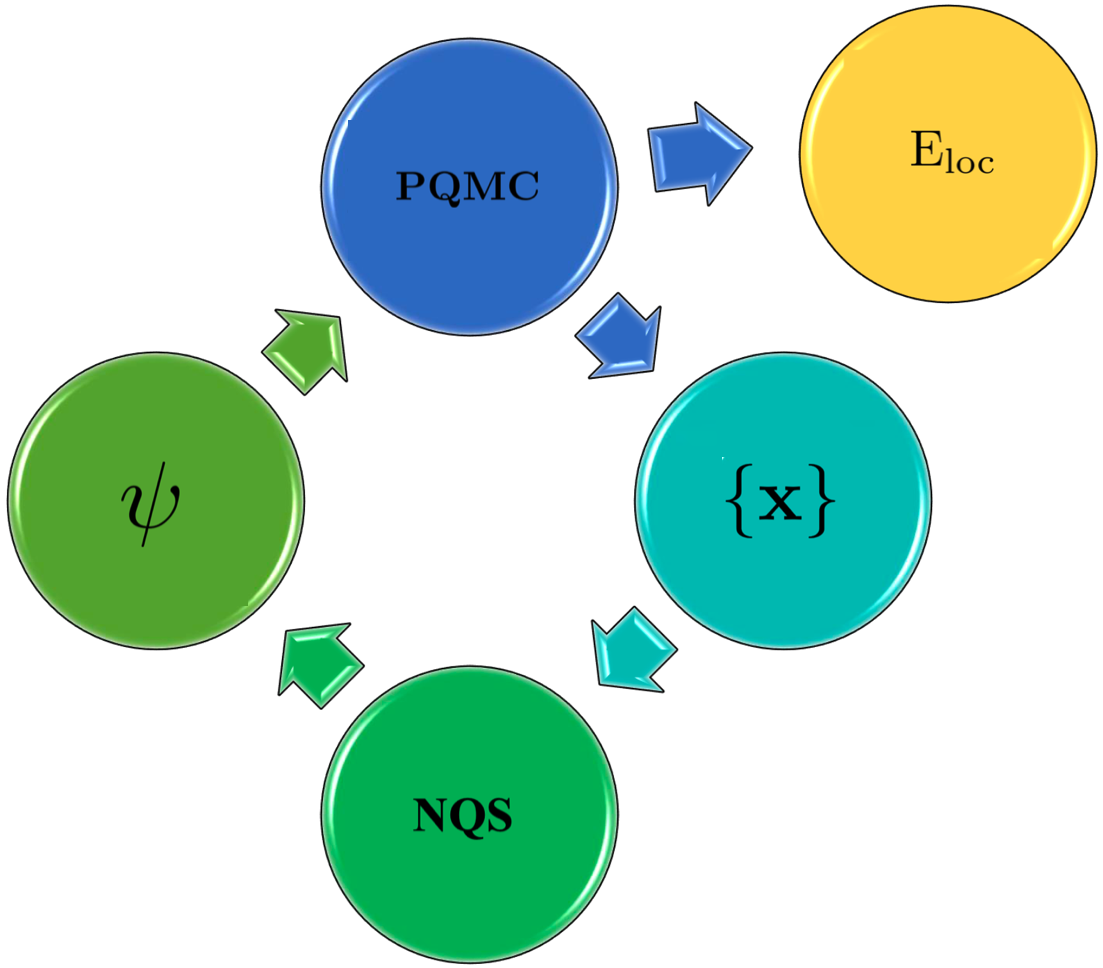}
    \caption{\textbf{PQMC guided by an NQS}. We represent a simplified schematic plot of the PQMC guided by an NQS. Here, the NQS uses the set of walkers $\{\textbf{x}$\} from the PQMC to train an ansatz $\psi$, which should be a ``good'' approximation of the target ground state. This ansatz is then used to guide the PQMC simulation and to compute the local energies $E_{\text{loc}}$. The ground state energy is obtained by averaging these local energies. }
    \label{F0}
\end{figure}

Since the advent of artificial intelligence (AI) and machine learning (ML), MC methods have assumed a central role in the approximation of the many-body wavefunction by neural network quantum states (NQSs)~\cite{wah2025many,hibat2020recurrent,pilati2019self,carleo2017solving,carrasquilla2021use,kim2024neural,wah2026bridging}. Among the plethora of MC methods, only VMC is currently commonly used in the context of NQSs, and its role is largely restricted to sampling configurations and stochastically computing the expectation values of observables~\cite{sorella2005wave}, thereby treating VMC as a complementary tool for NQSs. So far, very few works have proposed the reverse framework, namely using NQSs as a tool for QMC techniques~\cite{pilati2019self}. In this regard, we propose a framework [see Fig.~\ref{F0}] in which NQSs are used as a guiding tool for MC simulations. This technique has been employed using RBMs in Ref.~\cite{pilati2019self}. Here, we propose to extend this study to lattice models, and to an autoregressive PQMC algorithm in which the PQMC is guided by a RNN. We first employ it to determine the ground-state properties of Hermitian many-body systems. Furthermore, we extend our approach to non-Hermitian (NH) many-body systems--a regime where neither PQMC methods nor our autoregressive PQMC framework have been previously explored. This work is also motivated by the recent growing interest in the study of NH systems~\cite{wah2025many, solinas2025biorthogonal,wah2026bridging}.

The rest of the work is organized as follows. In Sec.~\ref{S1}, we present an overview of the PQMC methods we generalized in this work; in Sec.~\ref{S2}, we discuss our method of guiding PQMC with an RNN ansatz; in Sec.~\ref{S3}, we apply this method to a Hermitian model, and generalize to a non-Hermitian model in Sec.~\ref{S5}. Finally, we summarize our findings in Sec.~\ref{S4}.

\section{Projective Quantum Monte Carlo}\label{S1}

PQMC is a computational method that enables the simulation (at zero temperature) of the ground-state properties of quantum many-body systems described by stoquastic (sign-problem-free) Hamiltonians~\cite{PhysRevLett.74.3652, PhysRevLett.93.136405}. One of its key principles is the exponential suppression of higher-energy states of gapped Hamiltonians as time progresses. This suppression is achieved through the action of the projector operator, which projects out the ground state of the system during its imaginary-time evolution~\cite{inack2018phd}. We shall refer to this as the simple projective quantum Monte Carlo algorithm (SPQMC). While SPQMC simulations can provide an accurate estimate of the ground-state properties of a system, both the accuracy and efficiency can be further improved by employing importance sampling—that is, sampling only from the relevant configurations—and by guiding the walkers (equivalent copies of the system) toward the most relevant configuration space~\cite{pilati2019self} within the Hilbert space by introducing the so-called ``guiding or trial wavefunction,'' which we shall denote as $\psi_T$.

\subsection{Simple Projective Quantum Monte Carlo}

The adjective ``simple'' does not imply that the algorithm is trivial but rather that it corresponds to the standard form of PQMC, which does not require importance sampling (see Sec.~\ref{S1}.B) nor does it rely on a guiding wavefunction. Let $\bm{x}=\{x_0,x_1,...,x_N\}$ be the spin configuration of a given quantum system of $N$ spins with Hamiltonian operator $H$. PQMC simulations evolve a large number of replicas of the system (walkers) according to the imaginary Schr\"odinger equation
\begin{equation}\label{Eq1}
-\frac{d}{{d\tau}}\left| {\psi (\textbf{\textit{x}},\tau)} \right\rangle = ( {H}-E_{r})\left| {\psi (\textbf{\textit{x}},\tau)} \right\rangle,
\end{equation}
where $\tau$ is the imaginary time and $E_r$ is the reference energy that regulates the number of walkers by keeping it around a target (averaged) number as will be discussed later. The role of this energy will be discussed later. The similarity between Eq.~\eqref{Eq1} and the diffusion equation is the main reason why this method is often referred to as DMC. The iterative solution of Eq.~\eqref{Eq1} within a time step $\Delta\tau$ is given by
\begin{equation}\label{Eq2}
\psi (\bm{x},\tau+\Delta\tau) = \sum\limits_{x'} {G(\bm{x, x'},\Delta \tau )} \psi (\bm{x'},\tau ),
\end{equation}
where $\psi (\bm{x'},\tau )=\langle\bm{x'}|\psi(\tau)\rangle$, and $G(\bm{x}, \bm{x'},\Delta \tau )$ is the Green's function that physically represents the transition probability from a state (configuration) $\bm{x'}$ to a state $\bm{x}$ within an imaginary time interval $\Delta\tau$, such that
\begin{equation}\label{Eq3}
G(\bm{x}, \bm{x'},\Delta \tau )= \left\langle \textbf{\textit{x}} \right|{{\mathop{\rm e}\nolimits} ^{ - \Delta \tau ({H} - {E_{r}})}}\left| {\bm{x'}} \right\rangle.
\end{equation}

The Green's function (GF) above is not always normalized, which means that its matrix elements are not stochastic. We rewrite this Green's function in terms of an always stochastic Green's function $G_{S}(\bm{x}, \bm{x'},\Delta \tau )$, up to a normalization factor $b_{\bm{x'}}$ that depends on the current configuration $\bm{x'}$, and evolve many walkers through $G_{S}(\bm{x}, \bm{x'},\Delta \tau )$ such that
\begin{equation}\label{Eq4}
    G(\bm{x}, \bm{x'},\Delta \tau )=G_{S}(\bm{x}, \bm{x'},\Delta \tau )b_{\bm{x'}},
\end{equation}
where the normalization factor $b_{\bm{x'}}$ is the weight of the configuration $\bm{x'}$ given by
\begin{equation}\label{Eq5}
    b_{\bm{x'}}= \sum\limits_{\bm{x}} {G({\bm{x},\bm{x'}},\Delta \tau )}.
\end{equation}

\begin{figure}
\includegraphics[width=0.48\textwidth]{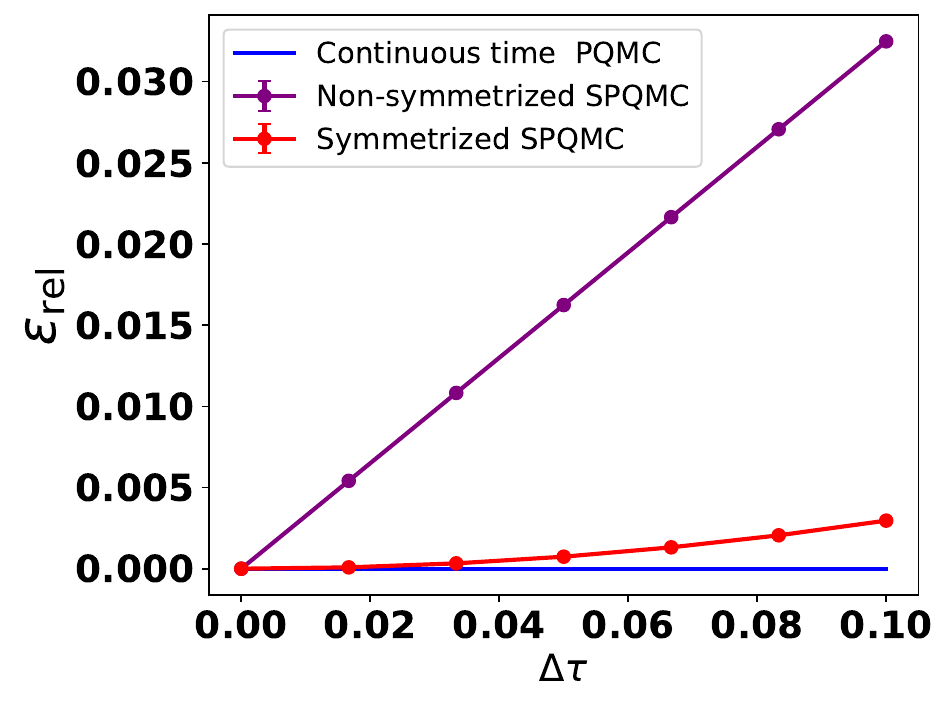}
\caption{\textbf{Bias due to time step}. The relative error $\varepsilon_{\text{rel}}$ is plotted against the imaginary time step $\Delta\tau$ for the continuous-time PQMC (blue), symmetrized SPQMC (red), and non-symmetrized SPQMC (purple). The simulation is performed for $N_w =10^3$ walkers, and $N=20$ spins. For continuous-time PQMC, a blue line is provided to guide the eye, as fitting is not meaningful in this case. The symmetrized SPQMC curve is fitted with a quadratic function $ax^2$, with optimal parameter $a=0.285(7)$, and the non-symmetrized SPQMC curve is fitted with a linear function $ax$, with optimal parameter $a=0.263(5)$. The Hermitian Ising chain in Eq.~\eqref{Eq9} is considered.}
\label{F1}
\end{figure}

While the system is normalized, since our study involves stochastic processes, not all walkers contribute equally. One may thus encounter a situation in which the contribution of some walkers becomes dominant, leading to large fluctuations in the walker contributions. This issue can be resolved by generating, for each walker in the population at a given time $\tau$--which we refer to as a ``father''--a number of corresponding ``sons'' in the population at $\tau+\Delta\tau$, such that the contribution of all walkers is properly accounted for. This ``death-birth'' procedure is called branching~\cite{PhysRevLett.74.3652, PhysRevLett.93.136405, inack2018phd}. Performing branching allows high-weight walkers to be split into several walkers, while low-weight walkers are either passed to the next generation or annihilated, resulting in fluctuations in the walker population at each PQMC iteration. These fluctuations can be stabilized by introducing a reference energy (referred to as a regulator), which keeps the number of walkers around a target value $N_{\text{avg}}$,
$E_r = E+\mu\log(N_{\text{avg}}/N_{\text{cur}})$,
where $N_{\text{cur}}$ is the size of the walker population at step $\tau$, $E$ is the averaged energy over the walker population $N_{\text{cur}}$ at the previous step (fathers), and $\mu$ is a small positive factor used to reduce the fluctuations. For a generic Hamiltonian $H=H_p+H_k$, where $H_p$ and $H_k$ are the potential and kinetic contributions, respectively (which do not commute in general), one can expand the propagator by means of the first-order Trotter approximation~\cite{inack2018phd,PhysRevB.33.6271}. However, this approximation introduces a bias of order $o(\Delta\tau^2)$ or $o(\Delta\tau^3)$ for the non-symmetrized or symmetrized approximation, respectively (see Fig.~\ref{F1}). We shall discuss later how to mitigate this bias. A crucial observation from all of the above is that the algorithm does not require any prior knowledge of the wavefunction, which makes SPQMC efficient. However, SPQMC stochastically explores the whole configuration space, and thus $H_p$ may vary greatly from one region to another and from one generation of walkers (fathers) to the next (sons). This can be mitigated through the so-called importance sampling.

\subsection{Importance Sampling}
In general, when direct sampling from a given distribution is not possible, one can instead sample from a distribution that is close to the original one. This is the idea of importance sampling—that is, sampling from the relevant distribution. Instead of sampling from the original distribution, one samples from a new distribution $f$ containing configurations with high probability amplitude, thereby guiding the simulation toward the region of configuration space, where these configurations contribute the most to the wavefunction. A guiding wavefunction can be used for this purpose~\cite{pilati2019self}, and the new distribution can be written as
\begin{equation}\label{Eq6}
f(\bm{x}, \tau) = \psi_{T}(\bm{x})\psi(\bm{x}, \tau).
\end{equation}
As $\tau \to \infty$, PQMC projects the wavefunction $\psi(\bm{x}, \tau)$ onto the ground state $\psi_0(\bm{x})$, so that $ f(\bm{x}, \tau\to\infty) = \psi_{T}(\bm{x})\psi_0(\bm{x})$. In addition, for $f$ to represent the probability of finding the system in its ground state, one requires $\psi_{T}(\bm{x})\approx\psi_0(\bm{x})$. As such, the only requirement imposed on the guiding wavefunction is that it should be a ``good'' approximation of the true ground-state wavefunction. This condition will be discussed in detail later.

Substituting Eq.~\eqref{Eq6} into Eq.~\eqref{Eq1} [see Appendix~\ref{app:der_greens_fct} for derivation], one can show that the Green's function in Eq.~\eqref{Eq3} becomes (in the presence of importance sampling)
\begin{equation}\label{Eq7}
\begin{array}{l}
\tilde G(\bm{x}, \bm{x'}, \Delta\tau) =\tilde G_{b}(\bm{x}, \bm{x'}, \Delta\tau) \tilde G_{d}(\bm{x}, \bm{x'}, \Delta\tau),
\end{array}
\end{equation}
where $\tilde G_{b}(\bm{x}, \bm{x'}, \Delta\tau)$ is the Green's function modeling the branching procedure and is a function of the local energy, and $\tilde G_{d}(\bm{x}, \bm{x'}, \Delta\tau)$ is the Green's function that defines the drift–diffusion process of walkers toward the relevant configuration space. The exact derivation of each of these Green's function is given in Appendix~\ref{app:der_greens_fct}. In this case, the local energy is constructed from a mixed estimator similarly as in Ref.~\cite{wah2025many}. The mixed estimator of any observable is computed during PQMC as
\begin{align}\label{Eq8}
\langle \mathcal{O}\rangle=\frac{\langle\psi_T(\bm{x})| \mathcal{O}|\psi(\bm{x},\tau)\rangle}{\langle\psi_T(\bm{x})|\psi(\bm{x},\tau)\rangle},
\end{align}
which is no longer upper bounded (variational). We shall discuss the consequences of this result later.

\begin{figure*}
    \centering
    \includegraphics[width=0.7\textwidth]{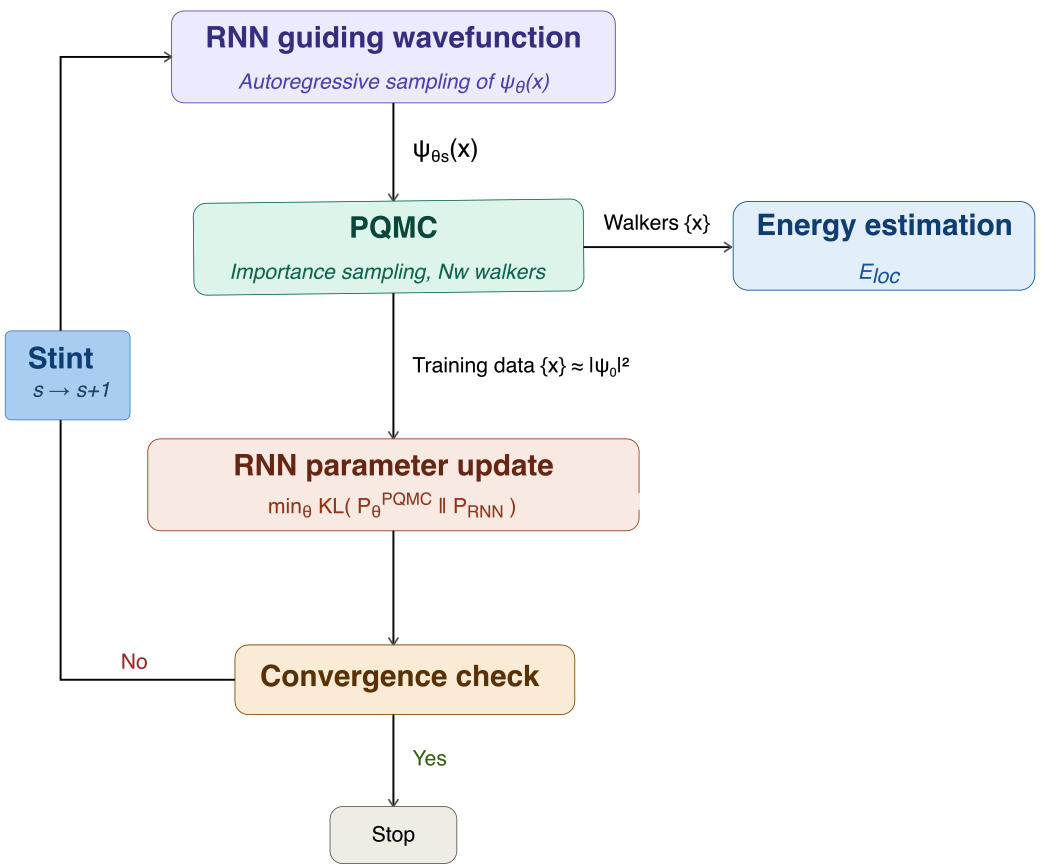}
    \caption{\textbf{Autoregressive PQMC}. We show a schematic plot of self-learning PQMC guided by an RNN. $\theta$ are the RNN parameters.  }
    \label{F2}
\end{figure*}

\subsection{Continuous Time Approximation}

First introduced in Refs.~\cite{sorella1998green,sorella2000green,becca2017quantum}, the idea behind the continuous-time approximation is to slice the MC time step $\Delta\tau$ into $m \to \infty$ small intervals, extract the time $\delta\tau_p$ that passes before the transition to the next configuration, and keep track of the remaining time $\delta\tau_r$ needed to complete the interval $\Delta\tau$ \cite{inack2018phd}. This process is iteratively performed while updating $\delta\tau_r \to \delta\tau_r - \delta\tau_p$ until $\delta\tau_r = 0$, at which point branching is performed using the total accumulated weight factor. This procedure is known as continuous-time PQMC. As shown in Fig.~\ref{F1}, employing this procedure mitigates the bias due to finite time steps. In this work we will use this continuous-time approximation to improve the accuracy of the PQMC.

\section{Autoregressive Projective Quantum Monte Carlo}\label{S2}
\subsection{Method}

PQMC algorithms are among the most powerful computational techniques for simulating the ground-state properties of quantum many-body systems. Their efficiency, however, depends critically on the use of a sufficiently accurate (``good'') trial wavefunction to guide the simulation. Usually, this guiding wavefunction is obtained from a separate variational optimization, as done with restricted Boltzmann machines (RBMs) in Ref.~\cite{pilati2019self}. Here, we propose using a recurrent neural network (RNN) as the guiding wavefunction. This approach offers several advantages: the autoregressive structure provides a direct factorization of the wavefunction (chain rule), allowing sequential sampling of configurations and yielding independent samples with tractable likelihoods~\cite{PhysRevLett.128.090501}. In contrast, RBMs (and unrestricted Boltzmann machines) require marginalizing over or sampling hidden units (via Gibbs sampling), introducing sampling overhead and correlations among samples, which necessitate block averaging or binning techniques~\cite{10.1007/978-3-642-15825-4_26}. Independent samples accelerate the PQMC training loop and improve the quality of the Kullback–Leibler (KL) \cite{kullback1951kullback} fit to walker distributions.
Furthermore, RNNs provide an explicit likelihood for each spin string, enabling maximum-likelihood (KL divergence) updates to be applied directly and stably using cross-entropy on PQMC walker data~\cite{jiao2024ai, PhysRevLett.124.020503}, which is numerically more stable than optimizing an RBM by minimizing KL divergence against PQMC samples~\cite{melko2019restricted, pilati2019self}. An RNN can also cheaply generate likely configurations (and conditional probabilities for partial flips), enabling fast, informed proposal moves (including block/sequential proposals) that reduce autocorrelation and branching noise~\cite{rautela2024conditional, kumar2025autoregressive,parthipan2023using}. 
Moreover, RNNs can straightforwardly parameterize both log-amplitude and phase (either via two networks or a complex output head) while maintaining the autoregressive likelihood, facilitating online learning of complex-valued guiding functions more easily~\cite{kumar2025autoregressive}. 

We employ an RNN as the guiding wavefunction within a self-learning PQMC framework. The PQMC algorithm maintains an ensemble of walkers ${\bm{x}}$, and at iteration $s$, the RNN provides an explicit, normalized probability amplitude $\psi_{\theta_s}(\bm{x})$ for any spin configuration. This autoregressive form allows exact evaluation of conditional probabilities and efficient independent sampling. At each stint $s$, PQMC is run for a fixed number of projection steps using $\psi_{\theta_s}(\bm{x})$ as the importance-sampling guiding function, producing an updated walker distribution approximating $|\psi_0|^2$. From this distribution, the local energy is computed. The set of walker configurations is then used to update the RNN parameters by maximizing the likelihood using the autoregressive factorization (equivalently, minimizing the Kullback–Leibler divergence between the RNN distribution and the PQMC walker distribution). This self-learning cycle [see Fig.~\ref{F2}] is repeated for stints $s=1,2,3,...$, with the guiding function becoming progressively more accurate [see Appendix~\ref{A1}], thereby reducing variance, improving importance sampling, and suppressing finite-walker population bias. Convergence is reached when $s \to \infty$, and successive stints yield stable RNN parameters and stationary PQMC energy estimates. Our results show that  $s=5$ stints suffice to reach the convergence [see Appendix~\ref{A1}].

\begin{figure}
\centering
\includegraphics[width=0.46\textwidth]{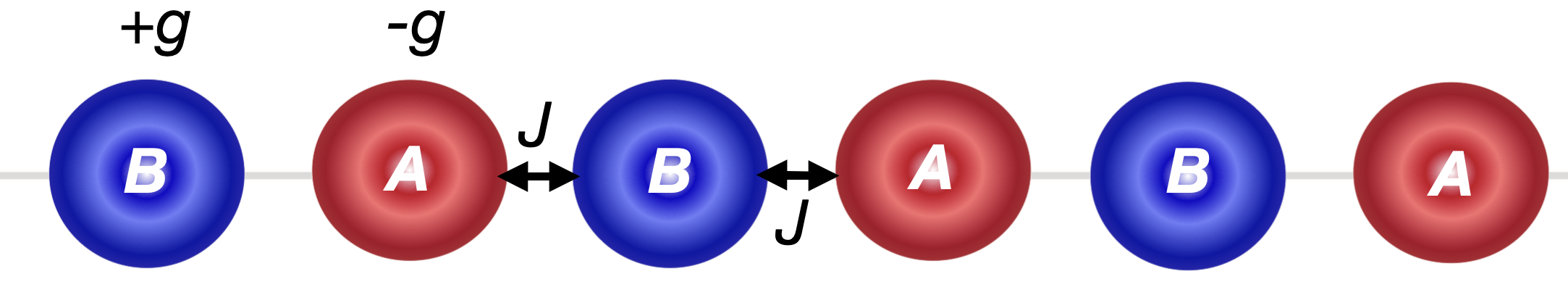}
\caption{\textbf{Hermitian 1D Ising chain with alternating magnetic field}. $A$ (red) and $B$ (blue) represent the sublattices in which spin-$1/2$ particles occupy the odd and even positions, respectively. The transverse field $+g$ ($-g$) is applied to the $B$ ($A$) sublattices. Nearest-neighbor coupling $J$ is considered.}
\label{FF}
\end{figure}

\subsection{Stoquastic Condition}

PQMC is generally well-suited for simulating the ground state of stoquastic Hamiltonians, i.e., Hamiltonians with real and non-positive off-diagonal elements. This stoquasticity ensures that the Hamiltonian is sign-problem-free~\cite{loh1990sign}. Even in the presence of a sign problem, some studies suggest that PQMC can still simulate the system, albeit at a high computational cost~\cite{pilati2019self,reynolds1982fixed}. Many efforts have been made to mitigate the sign problem, particularly for fermionic systems where the sign problem is usually severe~\cite{reynolds1982fixed,booth2009fermion,ceperley1984quantum,mori2017toward,li2015solving}. One of the simplest and most commonly used methods to mitigate this issue is a change of basis (which is not always possible)~\cite{li2015solving}.
In the context of NH systems, the sign problem becomes more pronounced because the off-diagonal elements of NH Hamiltonians are generally complex. In this work, we show that performing a gauge rotation [see Appendix~\ref{A2}] via a unitary transformation can help mitigate the sign problem for certain NH Hamiltonians. We note, however, that this method is not a general framework and may fail for arbitrary NH Hamiltonians.
Besides that, the sign problem can be mitigated in parity-time($PT$)-symmetric non-Hermitian Hamiltonians in the $PT$-unbroken regime by transforming them into Hermitian Hamiltonians via a similarity transformation~\cite{mostafazadeh2003exact,MIAO20161805}, as we will discuss later. This procedure introduces some nonlocality, however, as discussed in Appendix~\ref{A2}. For now, we will introduce our models. All parameters used for the RNN and PQMC simulations are summarized in Appendix~\ref{A3}. All experiments were performed on an HPC cluster using NVIDIA Quadro RTX 6000 GPUs (24\,GB VRAM, CUDA~12.2).

\section{Experiment on a Hermitian Model}\label{S3}
\subsection{Hamiltonian}

\begin{figure*}[ht!]
\centering
\subfloat[\centering ]{{\includegraphics[width=0.48\textwidth]{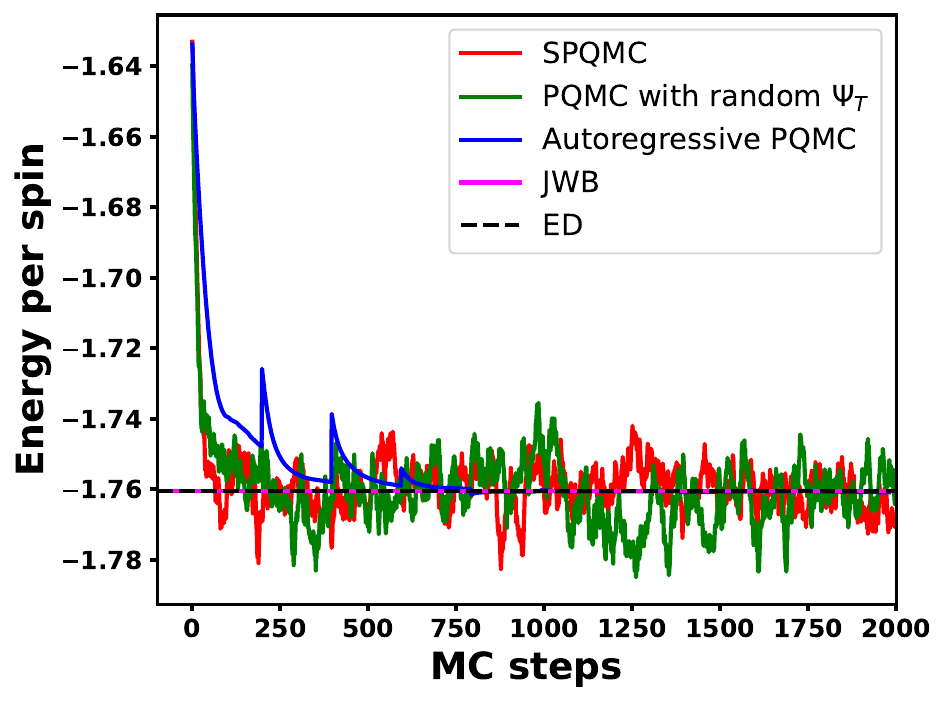} }}
\subfloat[\centering ]{{\includegraphics[width=0.48\textwidth]{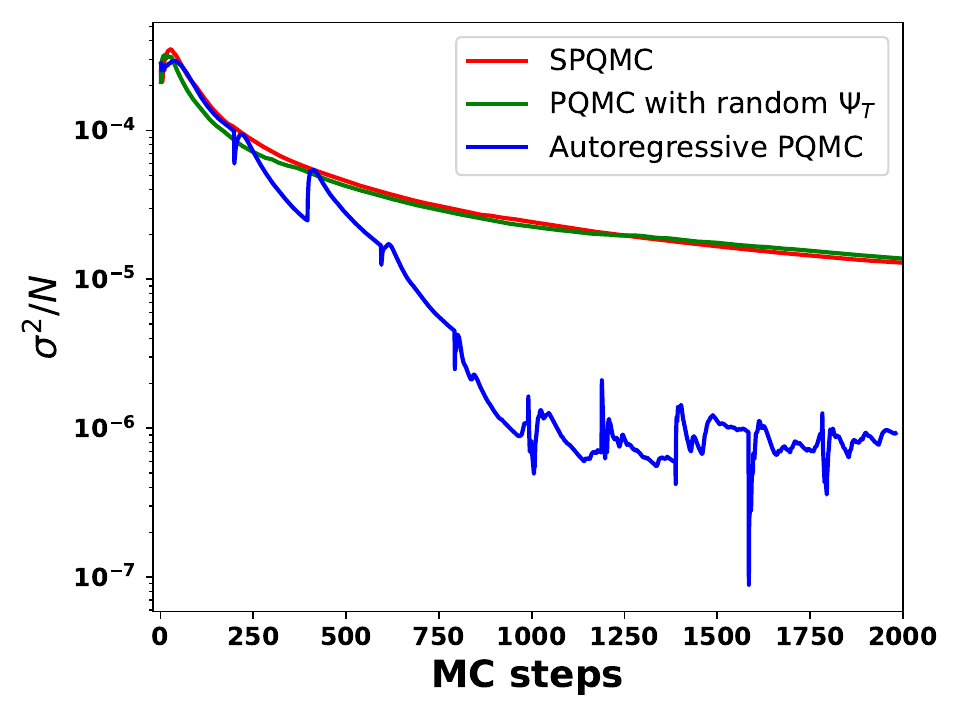} }}
\caption{\textbf{Autoregressive PQMC vs SPQMC and ED}. We plot \textbf{(a)} the energy per spin with SPQMC (red), and PQMC using a random $\psi_T$ (green) for $\Delta \tau = 10^{-2}$, $g=1.6$, $N_w=2\times10^4$, $MCs=2\times10^3$, and the projection time $\Delta \tau_p = MCs\times\Delta\tau=20$. The energy per spin obtained using autoregressive PQMC guided by the RNN ansatz is shown in blue for the same parameters with the number of stints $N_s=10$, number of MC iterations per stint $N_i=200$, number of training steps per stint $N_t=10^3$, and number of samples used to train the RNN $n_s=7\times10^2$. Results are benchmarked against ED (black dashed) and the analytical result (purple) obtained via the Jordan–Wigner–Bogoliubov (JWB) method. \textbf{(b)} The energy variance per spin for each method is shown. The guided PQMC (in blue) exhibits smaller fluctuations and its variance is much lower than that for SPQMC (red) or PQMC with a random trial wavefunction (green). Periodic boundary conditions (PBCs) are used and we considered $N=20$.}
\label{F3}
\end{figure*}

To benchmark our method, we first consider a 1D Hermitian transverse-field Ising model (TFIM) of $N$ spin-$1/2$ particles on a lattice coupled via a nearest-neighbor spin coupling $J$ [see Fig.~\ref{FF}]. The chain is subject to a staggered magnetic field, $+g$ on odd ($B$) sublattices and $-g$ on even ($A$) sublattices. The Hamiltonian of the system is given by
\begin{equation}\label{Eq9}
H = -J \sum_{\langle i,j \rangle} \sigma_i^z \sigma_j^z - g \sum_{i=1}^N (-1)^i \sigma_i^x,
\end{equation}
where $\sigma_j^z$ and $\sigma_j^x$ are Pauli matrices, and $\langle i,j \rangle$ denotes we sum over nearest-neighbor sites only. The choice of this model is motivated by the fact that it's a natural bridge to non-Hermitian physics. Alternating structure is the standard way to build a $PT$-symmetric lattice model. Aslo, the uniform TFIM's ground state is translationally invariant, so a sufficiently symmetric ansatz (or even weight-shared RNN cells) can exploit that structure fairly easily. Staggering the field forces the network to learn two inequivalent sublattices, that is a less trivial, lower-symmetry wavefunction. That makes it a better benchmark of the autoregressive sampler's expressivity than a model where a simple (symmetric) guess is close to correct. The Green's function in Eq.~\eqref{Eq7} for this model is derived in Appendix~\ref{app:der_greens_fct}.

\subsection{Exact ground state with Wigner-Jordan-Bogoliubov transformations }

The model in Eq.~\eqref{Eq9} can be exactly solved using the Jordan–Wigner–Bogoliubov (JWB) formalism~\cite{LIEB1961407} as follows. First, we remove the staggered field by a simple unitary transformation, let us define
\begin{equation*}
    U=\prod_{j\ \text{odd}}\sigma_j^z, \qquad U^\dagger=U,
\end{equation*}
such that the conjugation gives $U\sigma_i^z U=\sigma_i^z$ and $U\sigma_i^x U= (-1)^{i}\sigma_i^x$, therefore
\begin{equation*}
    H' \equiv U H U
= -J\sum_{i} \sigma_i^z \sigma_{i+1}^z \;-\; g\sum_{i} \sigma_i^x.
\end{equation*}
This suggests that the model is unitarily equivalent to the uniform-field TFIM. The ground state $|\Omega\rangle$ of $H$ is $|\Omega\rangle = U^\dagger |\Omega'\rangle = U |\Omega'\rangle$, where $|\Omega'\rangle$ is the ground state of $H'$.
Second, we rotate the axes to the standard TFIM form by applying a global $90^\circ$ rotation about $y$
\begin{align*}
    R=\prod_i e^{-i\frac{\pi}{4}\sigma_i^y}:\quad
R\sigma_j^z R^\dagger=\sigma_j^x,\quad
R\sigma_j^x R^\dagger=-\sigma_j^z,
\end{align*}
such that
\begin{align*}
    \widetilde H \equiv R H' R^\dagger
= -J \sum_i \sigma_i^x \sigma_{i+1}^x \;+\; g \sum_i \sigma_i^z.
\end{align*}

This is the standard quantum Ising chain (up to a harmless sign convention on $g$). After diagonalizing and performing the Jordan-Wigner mapping as well as the Bogoliubov transform, one obtains
\begin{align}\label{jwb}
    \frac{E_0(H)}{N} = -\frac{1}{\pi}\int_0^\pi \sqrt{(g - J\cos k)^2 + (J\sin k)^2}\,dk,
\end{align}
which represents the exact ground-state energy of our model [see Appendix~\ref{A5} for details].

\begin{table*}
\centering
\begin{tabular}{|c|c|c|c|c|c|}
\hline
Size & Method & $E_\textrm{loc}$  & $\sigma^2/N$ & $\varepsilon_\textrm{rel}$ & Time(hh:mm:ss) \\ \hline
 &JWB  & \textbf{-88.025406110236} & - & - & - \\
$N=50$ & SPQMC  & -88.035485019236 & $ 2.213 \times 10^{-4}$ & $ 1.145 \times 10^{-4}$ &01:37:11  \\
 &Autoregressive PQMC  & -88.025202331421 & $2.523 \times 10^{-5}$ & $2.315 \times 10^{-6}$ &01:50:37  \\
 \hline
 &JWB  & \textbf{-176.050812220459} & - & - & -  \\
$N=100$ & SPQMC  & -176.010647660133 & $ 2.284 \times 10^{-4}$ & $ 3.337 \times 10^{-4}$ &04:53:47 \\
 &Autoregressive PQMC  &  -176.049780386649 & $2.883 \times 10^{-5}$ & $5.861 \times 10^{-6}$ &07:22:23 \\
 \hline
 &JWB  & \textbf{-264.076218330689} & - & - & - \\
$N=150$ & SPQMC  & -264.378585600678 & $ 5.111 \times 10^{-4}$ & $ 1.141 \times 10^{-3}$ &07:14:59  \\
 &Autoregressive PQMC  & -264.073160328081 & $8.221 \times 10^{-5}$ & $ 1.158 \times 10^{-5}$ &10:00:01  \\
  \hline
\end{tabular}%
\caption{\textbf{Large system size simulations for Hermitian TFIM}. We compare our results from SPQMC and autoregressive PQMC for $N=50, 100, 150$ spins simulated with $N_w= 5 \times 10^4$, $N_w=10^5$, and $N_w= 1.5 \times 10^5$ walkers respectively. The energy variance per spin $\sigma^2/N$ and the relative error $\varepsilon_\textrm{rel}$ of the autoregressive PQMC with respect to the JWB method remain relatively low with increasing system size. The total projection time is the same for all the simulations.}
\label{T1}
\end{table*}

\begin{figure*}[ht!]
\centering
\subfloat[\centering ]{{\includegraphics[width=0.48\textwidth]{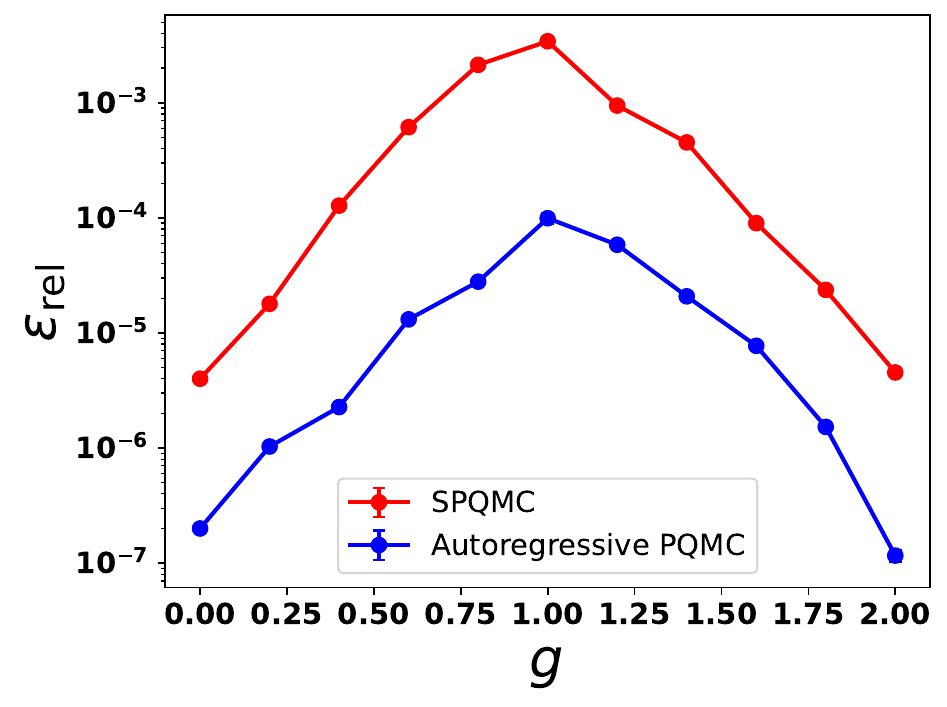} }}
\subfloat[\centering ]{{\includegraphics[width=0.48\textwidth]{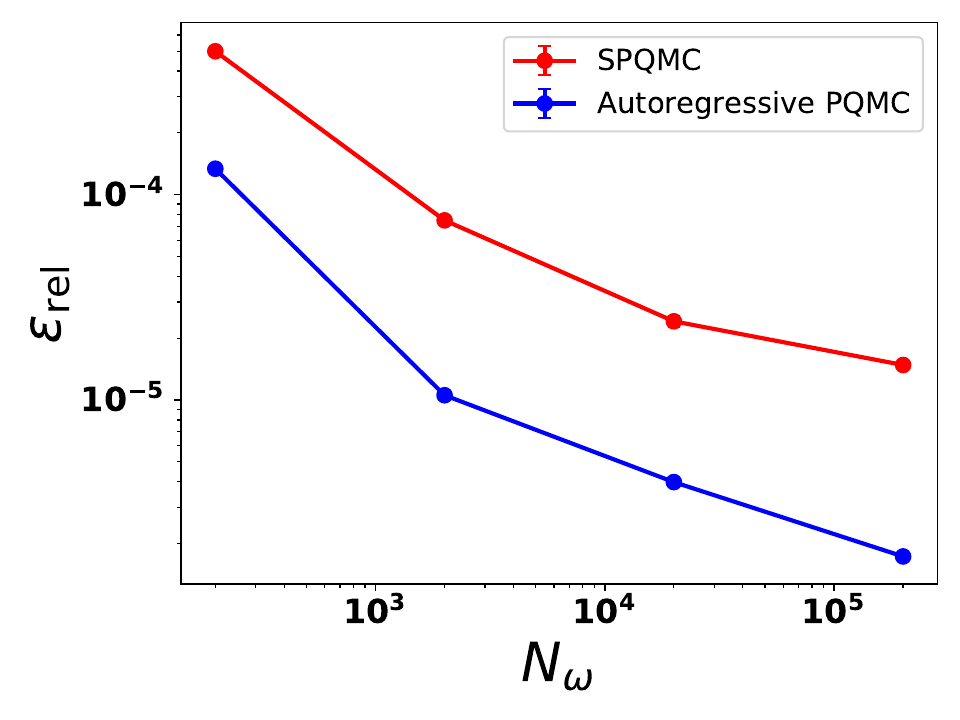} }}
\caption{\textbf{Relative errors}. We plot the relative error $\varepsilon_\textrm{rel}$ on the energy with respect to ED as a function of \textbf{(a)} the field strength $g$ using SPQMC (red) and neural-guided autoregressive PQMC (blue) for $\Delta \tau = 10^{-2}$, $N_w=2\times10^4$, $MCs=2\times10^3$, projection time $\Delta \tau_p = MCs\times\Delta\tau=20$, number of stints $N_s=10$, MC iterations per stint $N_i=200$, number of training steps per stint $N_t=10^3$, and number of samples used to train the RNN $n_s=7\times10^2$. \textbf{(b)} Relative error as a function of the number of walkers $N_w$ for $\Delta \tau = 10^{-2}$, $MCs=2\times10^3$, $\Delta \tau_p = MCs\times\Delta\tau=20$, $N_s=10$, and $g=1.6$. We assume $N=20$ and PBCs. Increasing the number of walkers improves the accuracy of both methods.}
\label{F4}
\end{figure*}

\subsection{Autoregressive Projective Quantum Monte Carlo}

We now implement the method described in the previous section. We perform $N_t=10^3$ unsupervised learning updates after each PQMC stint, observing convergence within $N_s=5$ to $N_s=10$ stints, cf. Fig.~\ref{F3}. The autoregressive nature of our ansatz makes this number of stints sufficient to approach the limit $s\to\infty$. The number of stints $N_s$ remain within this window regardless of the system's size, which is approximately four times less than the number of stints required for RBMs as reported in Ref.~\cite{pilati2019self} . 
In Fig.~\ref{F3}(a), we plot the ground-state energy per spin as a function of MC steps, and observe that the algorithm indeed converges to the true ground-state energy of the model. Unlike VMC, PQMC does not satisfy a variational principle. The energy is obtained from a mixed estimator (see Eq.~\eqref{Eq8}), which is not constrained to lie above the true ground-state energy. Consequently, instantaneous values of the local energies may exhibit stochastic fluctuations both above and below the true ground state. These excursions are amplified by several well-known sources of noise: the finite walker population used for branching, the intrinsic variance of the local-energy estimator, the quality of the guiding wavefunction, and discretization errors in the short-time propagator. Together or individually, these effects can temporarily bias the mixed estimator downward, even though the underlying projected state remains physically correct. Importantly, only the long-time average of the local energies is guaranteed to converge to the true ground state; individual samples are not variationally bounded.
A key consequence of the above observation is that the more expressive the ansatz, the smaller the fluctuations and the more accurate (``variational-like'') PQMC becomes. This is evident in Fig.~\ref{F3}(a), where our method exhibits significantly reduced fluctuations around the true ground state (blue curve) compared to SPQMC (red curve) or PQMC with a random ansatz (green curve). This indicates that our mixed estimator is nearly exact, i.e., our ansatz is a ``good'' approximation of the true ground-state wavefunction, such that

\begin{figure}
\centering
\includegraphics[width=0.5\textwidth]{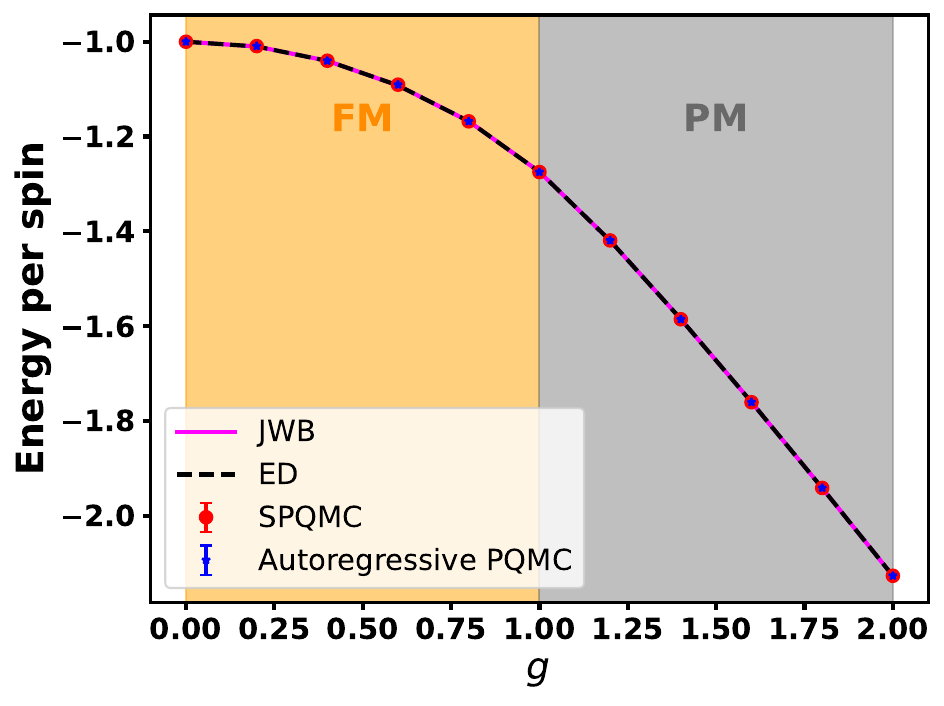}
\caption{\textbf{Phase diagram}. Ground-state energy per spin for both the ferromagnetic~(FM, orange background) and paramagnetic~(PM, gray) regimes for $N=20$ spins. Both methods are in good agreement with ED and JWB methods. PBCs are considered.}
\label{F5}
\end{figure}

\begin{align}\label{Eq26}
\frac{\langle\psi_T(\bm{x})| \mathcal{O}|\psi(\bm{x},\tau \to \infty)\rangle}{\langle\psi_T(\bm{x})|\psi(\bm{x},\tau \to \infty)\rangle} \approx \frac{\langle\psi_0(\bm{x})| \mathcal{O}|\psi_0(\bm{x})\rangle}{\langle\psi_0(\bm{x})|\psi_0(\bm{x})\rangle}.
\end{align}
Each peak in Fig.~\ref{F3}(a) marks the beginning of a new stint, showing that the algorithm becomes progressively more accurate with each stint.
Fig.~\ref{F3}(b) presents the energy variance, demonstrating that autoregressive PQMC (in blue) converges to a lower variance than SPQMC (red) or PQMC with a random trial wavefunction (green).

We now introduce the definition of the local energies, which are used to compute (by averaging) the ground-state energy 
\begin{align}\label{Eq27}
    E_{\text{loc}}(\bm{x})=E_p(\bm{x})-g\frac{\sum_{i=1}^N\psi_T(\bm{\tilde{x}_i})}{\psi_T(\bm{x_i})},
\end{align}
where $E_p$ denotes the diagonal (potential energy) contribution of the Hamiltonian in the computational basis, and $|\bm{\tilde{x}_i}\rangle =\sigma_i^x|\bm{x_i}\rangle$, is the updated spin configuration with the $i^{th}$ spin flipped. The probability for proposing such an update for SPMQC is $1/N$ if the previous spin configuration $\bm{x'}$ differs from the current spin configuration $\bm{x}$ by a single flip. Likewise for the autoregressive PQMC this probability is $\psi_T(\bm{x_i})/\sum_{i=1}^N\psi_T(\bm{\tilde{x}_i})$. Note that SPQMC can be obtained from the autoregressive PQMC by taking a constant (real) trial wavefunction reducing the local energy to $E_{\text{loc}}(\bm{x})=E_p(\bm{x})-gN$. We show in Table~\ref{T1} that our method remains robust and can be used to accurately simulate large system size within a reasonable amount of time. We stress the fact that, in order to make the ansatz as expressive as possible one needs to include more walkers as the system size increases, cf. Table~\ref{T1}. However, while changing the number of walkers $N_w$ [see Table~\ref{T1}] it is important to adjust the MC time step $\Delta \tau$ such that the total projection time $\Delta \tau_p = MCs\times\Delta\tau=20$ remains the same for all simulations.

Further, in Fig.~\ref{F4}(a), we compute the relative error in the energy per spin with respect to ED for different values of the transverse field $g$. In general, the relative error increases in the ferromagnetic ($g<1$) phase, decreases in the paramagnetic ($g>1$) phase, and reaches its maximum near the critical point ($g_c\approx1$), which is generally more challenging to capture. Notably, for $N=20$ spins, the autoregressive PQMC is at least one order of magnitude more accurate than SPQMC. We also show that the accuracy of both methods can be improved by increasing the number of walkers, cf. Fig.~\ref{F4}(b), although this comes at the cost of higher computational effort. Additionally, we plot the phase diagram of our model for $N=20$ spins, demonstrating good agreement between ED and JWB methods [see Fig.~\ref{F5}].

For completeness, we further demonstrate the effectiveness of our method by comparing it with variational NQSs~\cite{hibat2020recurrent,wah2025many} in Appendix~\ref{A4}. As expected, our method outperforms variational RNNs.

In the first part of this section, we showed the application of our method to a many-body Hermitian system. We demonstrated that autoregressive PQMC is a powerful tool for capturing ground-state properties, outperforming its vanilla version (SPQMC). Having established that PQMC guided by neural quantum states surpasses simple PQMC, we now pursue a different goal: extending the method to non-Hermitian systems. We will show that it also outperforms traditional variational NQS approaches (VMC + NQSs), using an RNN as the NQS. 

\begin{figure}
    \centering
    \includegraphics[width=0.46\textwidth]{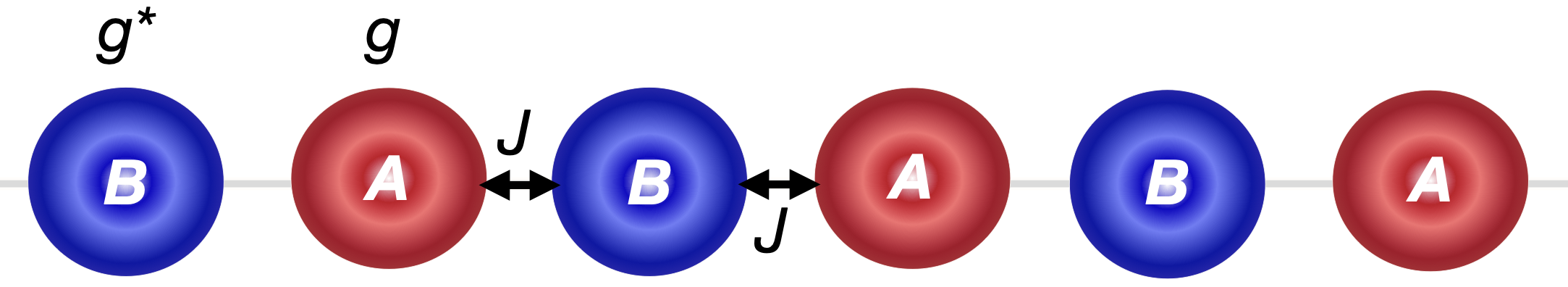}
    \caption{\textbf{Non-Hermitian 1D Ising chain in a staggered magnetic field}. $A$ (red) and $B$ (blue) represent the sublattices in which spin-$1/2$ particles occupy the odd and even positions, respectively. The complex field $g$ ($g^*$) is applied to the $A$ ($B$) sublattices~\cite{wah2025many}. Nearest-neighbor coupling $J$ is considered. }
    \label{F6}
\end{figure}

\section{Experiment on a Non-Hermitian Model }\label{S5}

A fundamental principle of quantum mechanics states that observables—such as the Hamiltonian of a closed system—must be self-adjoint, and are represented by Hermitian matrices~\cite{RevModPhys.93.015005}. In reality, physical systems are never completely isolated and always interact to some degree with their environment, introducing dissipative effects that complicate their descriptions. Non-Hermitian quantum systems have emerged as a powerful framework for modeling such scenarios. In NH systems, the Hamiltonian spectrum is generally complex (unless constrained by a symmetry), making the extension of the method presented here non-trivial, as PQMC simulations are prone to severe sign problems, as we will demonstrate in the following. In addition, to date, NH systems have only been studied with variational NQSs~\cite{wah2025many,solinas2025biorthogonal, wah2026bridging}, and neither simple PQMC nor PQMC guided by an NQS have been used to study such systems. This section is devoted to the filling of this gap. 

\begin{figure*}[ht!]
\centering
\subfloat[\centering ]{{\includegraphics[width=0.48\textwidth]{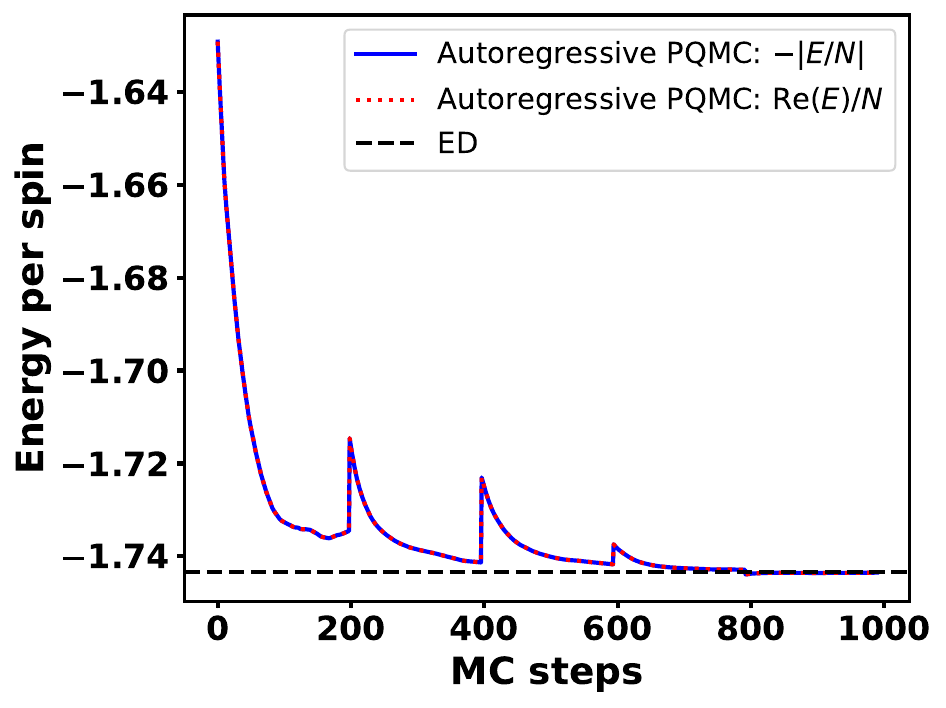} }}
\subfloat[\centering ]{{\includegraphics[width=0.48\textwidth]{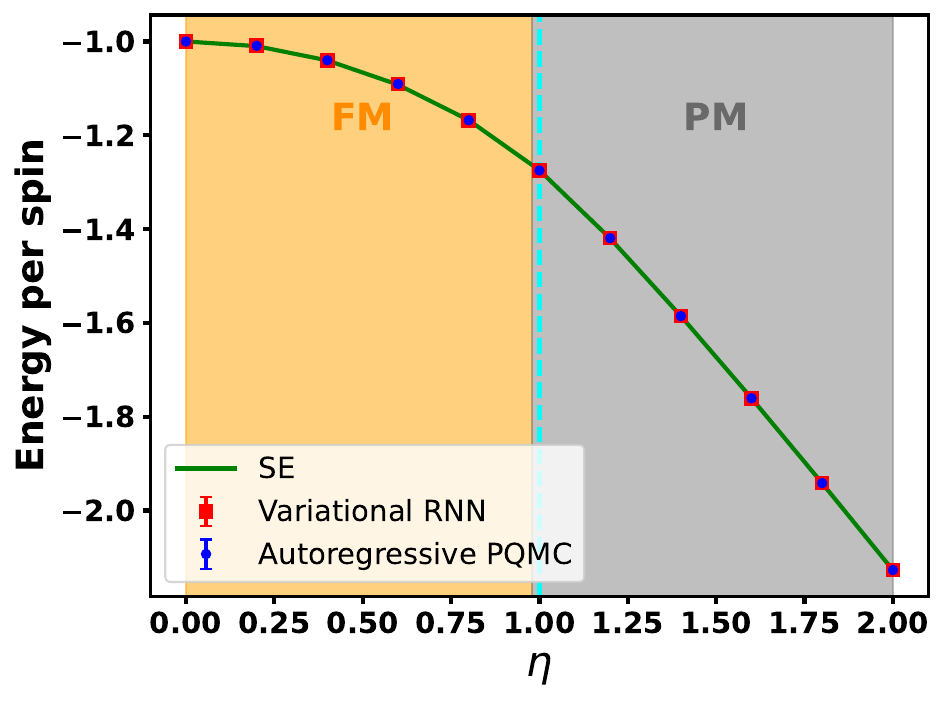} }}
\caption{\textbf{Energy per spin}. We compute the \textbf{(a)} real (red dashed) and absolute value (blue) of the ground-state energy per spin using autoregressive PQMC and benchmark against ED (black dashed) for $N=10$. \textbf{(b)} Phase diagram: energy per spin as a function of the real part of the field for $N=100$, comparing variational RNN (red) from Ref.~\cite{wah2025many} with autoregressive PQMC (blue), benchmarked against high-order series expansions~\cite{wah2025many,PhysRevB.104.195137}. The doted line (cyan) represent the transition (critical) line of the previous Hermitian model, we observe that the introduction of non-Hermiticity has the effect of shrinking the ferromagnetic (FM) phase. The gray region is the paramagnetic (PM) phase. OBCs are used.}
\label{F7}
\end{figure*}

\subsection{Hamiltonian and non-Hermitian symmetry}

We consider a one-dimensional (1D) parity-time-symmetric transverse field Ising chain of $N$ spin-$1/2$ particles on lattice. The transverse staggered magnetic field is chosen to be complex $g= \eta + i \xi$ (applied on sublattice $A$ (red)) and $g^*= \eta - i \xi$ (applied on sublattice $B$ (blue)), with $\eta, \xi \in \mathbb{R}$. When $\xi\neq0$, the system is thus non-Hermitian [see Fig.~\ref{F6}], The Hamiltonian representing this setting is~\cite{wah2025many,PhysRevB.104.195137}
\begin{align}\label{Eq28}
    H = -J \sum_{j=1}^N \sigma_j^z \sigma_{j+1}^z -g \sum_{j \in A}\sigma_j^x-g^* \sum_{j \in B}\sigma_j^x,
\end{align}
where the parity operator $P$ acts as $P \sigma_j^{\nu} P^{-1} = \sigma_{N-j+1}^{\nu}$, the time-reversal operator $T$ acts as $T i T^{-1} = -i$, and $T\sigma_j^{\nu} T^{-1} = \sigma^{\nu}_j$ with $\nu =x,z$. For convenience, unless stated otherwise, we shall work in the $PT$-unbroken regime, where the eigenvalues are real~\cite{bender2005introduction}. 

The NH Hamiltonian in Eq.~\eqref{Eq28} is not stoquastic, as its off-diagonal elements (the $g$ term) are complex. Consequently, this Hamiltonian is not sign-problem-free. While PQMC can simulate such a system (at least for small system sizes) despite the sign problem, we demonstrate in Appendix~\ref{A2} that the sign problem can be mitigated by performing a unitary transformation $U$ that rotates the basis. The unitary condition ensures that the spectrum of the system remains unchanged under this transformation, while effectively removing the phase factor in the field. Although this method works for the model considered here, it is important to note that it cannot be generalized to all NH Hamiltonians; nevertheless, it provides a useful starting point for mitigating the sign problem in NH systems, and generalizations to other NH systems can be an interesting avenue for future work. After this transformation, one can apply the prescription described in Sec.~\ref{S2} to simulate the ground-state properties of the NH model.

\subsection{Autoregressive Projective Quantum Monte Carlo}

\begin{figure}
\centering
\includegraphics[width=0.49\textwidth]{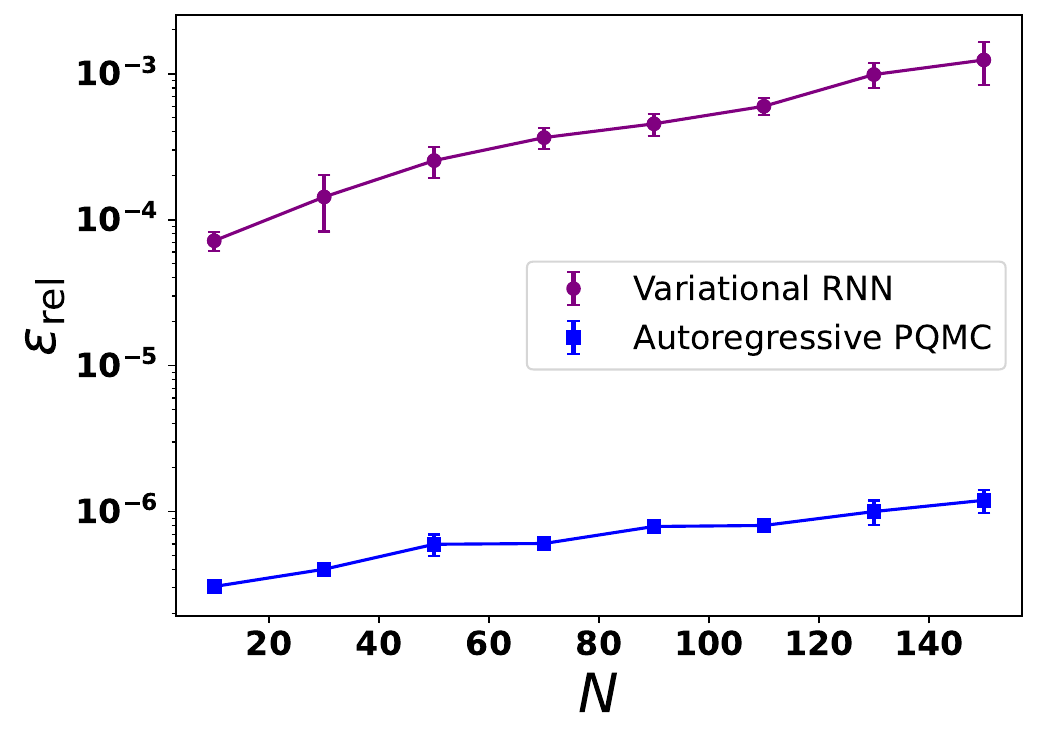}
\caption{\textbf{Relative error}. Relative error achieved with variational RNN (purple) and autoregressive PQMC (blue) as a function of system size. OBCs are considered. The autoregressive PQMC exhibits slower scaling and a lower relative error in the energy with increasing system size.}
\label{F8}
\end{figure}

In Fig.~\ref{F7}, we simulate the ground-state energy per spin of the NH Ising chain in Eq.~\eqref{Eq28}. We observe that the ground-state energy are in the spontaneously $PT$-unbroken regime, as shown in Fig.~\ref{F7}(a). Specifically, under open boundary conditions (OBCs) and for $g=1.6$, the real part of the energy (red) is equivalent to the negative absolute value of the energy, with the negative sign added for convenience. Within this parameter regime, all local energies are thus real. Accordingly, the autoregressive PQMC captures the symmetry of the model and converges to its true ground state. In this work, we define the ground-state energy as the state with the smallest real part of the energy.
In a previous work, we demonstrated that variational RNNs could simulate the ground state of this model for even large system sizes~\cite{wah2025many}. Here, we compare the variational RNN results to the autoregressive PQMC and show that, for $N=100$ [see Fig.~\ref{F7}(b)], the autoregressive PQMC accurately captures the ground-state properties of the model. While both methods converge to the true ground state, the relative error with respect to a higher-order series expansions~\cite{PhysRevB.104.195137} achieved by the autoregressive PQMC is smaller and exhibits slower scaling with system size [see Fig.~\ref{F8}].

For the non-Hermitian model, the local energy is calculated using~\cite{wah2025many}
\begin{equation}\label{Eq29}
E_\textrm{loc}(\bm{x}) = \frac{\langle \bm{x} | H | \Psi_{0,R} \rangle}{\langle \bm{x} | \Psi_{0,R} \rangle},
\end{equation}
where $|\Psi_{0,R}\rangle$ denotes the right ground-state wavefunction of $H$, which satisfies the biorthonormal condition
\begin{equation}\label{Eq30}
\langle \Psi_{i,L} | \Psi_{j,R} \rangle = \delta_{i,j}
\end{equation}
with $|\Psi_{0,L}\rangle$ the left ground-state wavefunction. Here we use that in the biorthogonal formalism, the expectation value of a non-Hermitian observable $\mathcal{O}$ is given by
\begin{equation}\label{Eq31}
\langle \mathcal{O} \rangle = \frac{\langle \Psi_{0,L} | \mathcal{O} | \Psi_{0,R} \rangle}{\langle \Psi_{0,L} | \Psi_{0,R} \rangle}.
\end{equation}
Since the model is $PT$-symmetric, one can show that $\langle \Psi_{i,L}| = (|\Psi_{i,R}\rangle)^\dagger$. Therefore, the biorthogonal product simplifies to a mixed estimator as for Hermitian systems. Additionally, Eq.~\eqref{Eq29} offers the practical advantage that computing the local energies does not require both the left and right eigenstates of the Hamiltonian, which would generally necessitate using two separate NQSs~\cite{wah2026bridging}.

\section{Conclusions and outlooks}\label{S4}

In this work, we introduce an autoregressive projective quantum Monte Carlo (PQMC) framework, in which a recurrent neural network (RNN) guides stochastic dynamics towards the relevant configuration space for determining the ground-state properties of many-body systems. Applied to both Hermitian and non-Hermitian 1D Ising chains, our method consistently offers several advantages over using a variational RNN, and outperforms standard PQMC. Beyond achieving higher accuracy and fidelity, this work establishes two previously unexplored firsts: the integration of autoregressive neural sampling into PQMC, and its application to discrete lattice models beyond continuum systems and to non-Hermitian systems. By providing a pathway to mitigate the sign problem, our framework opens the door to scalable simulations of low-energy states in complex quantum many-body systems, paving the way for future explorations in higher dimensions and across arbitrary non-Hermitian Hamiltonians.

\section{Acknowledgments}

L.W. gratefully acknowledges Estelle Inack and Roger Melko, whose insights and discussions during a previous project were extremely useful for this work. We thank Kai Phillip Schmidt, Jan Alexander Koziol, Lea Lenke, and Calvin Kr\"amer for insightful discussions on the model Hamiltonian, the exact diagonalization implementation and the discussion on the local energy. L.W. and F.K.K. acknowledge funding from the Max Planck Society Lise Meitner Excellence Program~\mbox{2.0}. F.K.K. also acknowledges funding from the European Union's ERC Starting Grant ``NTopQuant'' (101116680). Views and opinions expressed are however those of the authors only and do not necessarily reflect those of the European Union or the European Research Council (ERC). Neither the European Union nor the granting authority can be held responsible for them. Part of this work was carried out while R.Z. was affiliated with Max Planck Institute for the Science of Light.

\section{Data Availability}

The data and codes that support the findings of this study are available from the corresponding author upon reasonable request.

\appendix

\section{Derivation of the Green's function}\label{app:der_greens_fct}

We start by writing the Schr\"odinger equation in the conventional Dirac notation
\begin{equation}\label{Eq1.1}
 -i\hbar \frac{d}{{dt}}\left| {\psi (\textbf{\textit{x}},t)} \right\rangle  = (\hat{H}-E_{r})\left| {\psi (\textbf{\textit{x}},t)} \right\rangle,
\end{equation}
By performing a Wick rotation one obtains the so-called imaginary time Schr\"odinger equation
\begin{equation}\label{Eq1.2}
  -\frac{d}{{d\tau}}\left| {\psi (\textbf{\textit{x}},\tau)} \right\rangle  = ( \hat{H}-E_{r})\left| {\psi (\textbf{\textit{x}},\tau)} \right\rangle,
\end{equation}
where we use Gaussian units ($\hbar=1$). The iterative solution to this equation reads
\begin{equation}\label{Eq1.3}
  \psi (\bm{x},\tau+\Delta\tau) = \sum\limits_{x'} {G(\bm{x, x'},\Delta \tau )} \psi (\bm{x'},\tau ),
\end{equation}
where $G(\bm{x, x'},\Delta \tau )$ is the Green's function that physically represents the transition probability from a configuration $\bm{x'} $ to a configuration $\bm{x} $ within an imaginary time interval $\Delta\tau$ such that
\begin{equation}\label{Eq1.4}
  G(\bm{x, x'},\Delta \tau )= \left\langle \textbf{\textit{x}} \right|{{\mathop{\rm e}\nolimits} ^{ - \Delta \tau ({H} - {E_{r}})}}\left| {\bm{x'}} \right\rangle, 
\end{equation}
and 
\begin{equation}\label{Eq1.5} 
  \psi (\textbf{\textit{x}},\tau ) = \left\langle {\textit{\textbf{x}}\,|\psi (\textbf{\textit{x}},\tau )} \right\rangle.  \notag
\end{equation}
We normalized the above Green's function, such that
\begin{equation}\label{Eq1.6}
    G(\bm{x, x'},\Delta \tau )=G_{T}(\bm{x, x'},\Delta \tau )b_{\bm{x'}}, \notag
\end{equation}
where the normalization factor $b_{\bm{x'}}$ is the weight of the configuration $\bm{x'}$ given by
\begin{equation}\label{Eq1.7}
    b_{\bm{x'}}= \sum\limits_{\bm{x}} {G({\bm{x,x'}},\Delta \tau )}. \notag
\end{equation}

From Eq.~\eqref{Eq1.4} the Green's function for the 1D TFIM in Eq.~\eqref{Eq9} in the case of the non-symmetrized Trotter approximation is
\begin{equation}\label{Eq1.44}
 G \left( {{\bm{x},\bm{x'}},\Delta \tau } \right) = \left\langle \bm{x} \right|{e^{ - \Delta \tau {{ H}_{1}}}}{e^{ - \Delta \tau \left( {{{ H}_{2}} - {E_{r}}} \right)}} + \mathcal{O}\left( {\Delta {\tau ^2}} \right)\left| {x'} \right\rangle,
\end{equation}
where ${ H}_{1}$ is the transverse field term responsible for quantum fluctuations, and ${H}_{2}$ is the interaction term, which is diagonal in the basis state $\left| \bm{x} \right\rangle$. As such we find
\begin{widetext}
\begin{align} 
G\left({{\bm{x},\bm{x'}},\Delta \tau } \right) &= \left\langle \bm{x} \right|{e^{ - \Delta \tau {{\hat H}_{kin}}}}\left| \bm{x'} \right\rangle {e^{ - \Delta \tau \left( {{{ E}_{cl}} - {E_{r}}} \right)}} + \mathcal{O}\left( {\Delta {\tau ^2}} \right), \nonumber \\
&= \left\langle {{x_1}...{x_N}} \right|{e^{ - \Delta \tau \left( {\sigma _1^x...\sigma _N^x} \right)\Gamma }}\left| {{{x_1}'}...{{x_2}'}} \right\rangle {e^{ - \Delta \tau \left( {{{E}_{cl}} - {E_{r}}} \right)}} + \mathcal{O}\left( {\Delta {\tau ^2}} \right), \nonumber \\
&= \prod\limits_{i = 1}^N {\left\langle {{x_i}} \right|{e^{ - \Delta \tau \sigma _i^x\Gamma }}\left| {{{x_i}'}} \right\rangle } {e^{ - \Delta \tau \left( {{{E}_{cl}} - {E_{r}}} \right)}} + \mathcal{O}\left( {\Delta {\tau ^2}} \right). \label{Eq1.45}
\end{align}
After some calculations and assuming that only $\delta$ spins of a total of $N$ spins are flipped within $\Delta \tau$ we obtain
\begin{align}
G\left( {{\bm{x},\bm{x'}},\Delta \tau } \right) &= {c^N}{e^{ - \Delta \tau {b }\delta }}{e^{ - \Delta \tau {b }(N - \delta )}}{{\rm{e}}^{ - \Delta \tau ({E_{cl}}\left( \bm{x'} \right) - {E_{r}})}} + \mathcal{O}(\Delta {\tau ^2}), \nonumber\\
 & \simeq {\left[ {\cosh\Delta \tau \Gamma } \right]^N}{\left[ {\tanh \Delta \tau \Gamma } \right]^\delta }{{\rm{e}}^{ - \Delta \tau ({E_{cl}}\left( \bm{x'} \right) - {E_{r}})}}, \label{Eq1.48}
\end{align}
where $c = \sqrt {\frac{{\sinh 2\alpha}}{2}} $ and ${b } =  - \frac{1}{2}\log \tanh \alpha$.
Next, we normalize the Green's function and find we can separate it into two contributions
\begin{align}
    {G}\left( {{\bm{x},\bm{x'}},\Delta \tau } \right)={G_d}\left( {{\bm{x},\bm{x'}},\Delta \tau } \right) {G_b}\left( {{\bm{x},\bm{x'}},\Delta \tau } \right)
\end{align}
with
\begin{equation}\label{Eq1.49}
    \begin{array}{l}
{G_b}\left( {{\bm{x},\bm{x'}},\Delta \tau } \right) = {{\rm{e}}^{ - \Delta \tau ({E_{cl}}\left( {x'} \right) - {E_{r}})}}\sum\limits_x {{{\left[ {\cosh \Delta \tau \Gamma } \right]}^N}{{\left[ {\tanh \Delta \tau \Gamma } \right]}^\delta }}, \\
{G_d}\left( {{\bm{x},\bm{x'}},\Delta \tau } \right) = \frac{{G\left( {x,x',\Delta \tau } \right)}}{{\sum\limits_x {G\left( {x,x',\Delta \tau } \right)} }} = P_f^\delta {\left( {1 - {P_f}} \right)^{N - \delta }}.
\end{array}
\end{equation}
\end{widetext}
Here, ${G_b}\left( {{\bm{x},\bm{x'}},\Delta \tau } \right)$ is the Green's functions portraying the branching, and ${G_d}\left( {{\bm{x},\bm{x'}},\Delta \tau } \right)$ is the Green's function portraying the diffusion process of spin configurations.
${P_f} = \frac{{\sinh \Delta \tau \Gamma }}{{{e^{\Delta \tau \Gamma }}}}$ is the single-spin flip probability, which is performed by either flipping every spin, or alternatively by randomly selecting $\delta$-spins to be flipped from a binomial probability distribution, then randomly and uniformly select which spins to flip.

\begin{figure*}[ht!]
    \centering
    \subfloat[\centering ]{{\includegraphics[width=0.48\textwidth]{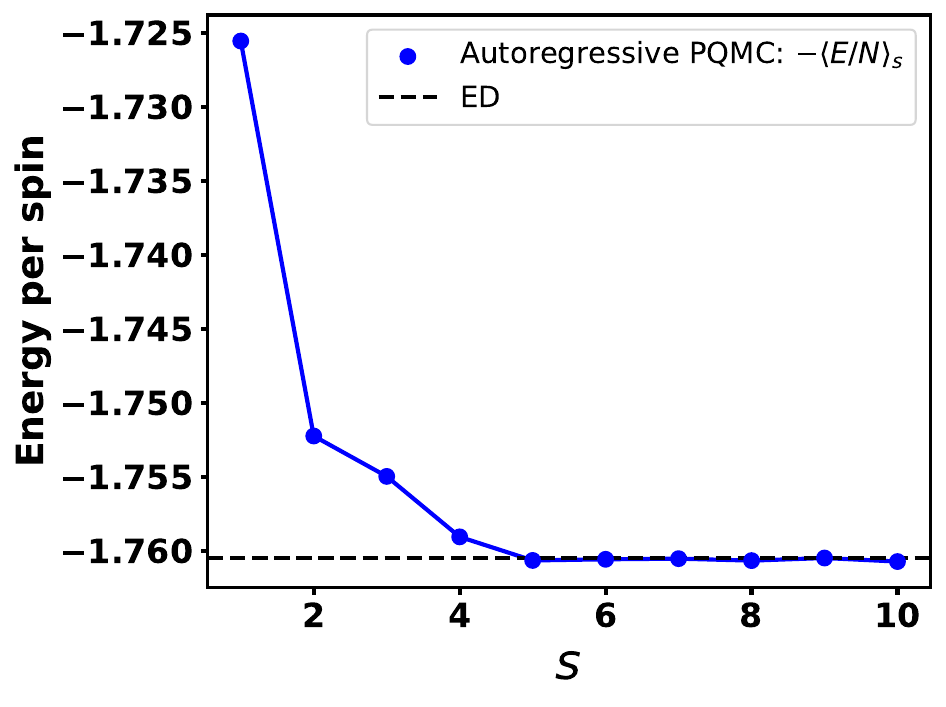} }}
    \subfloat[\centering ]{{\includegraphics[width=0.48\textwidth]{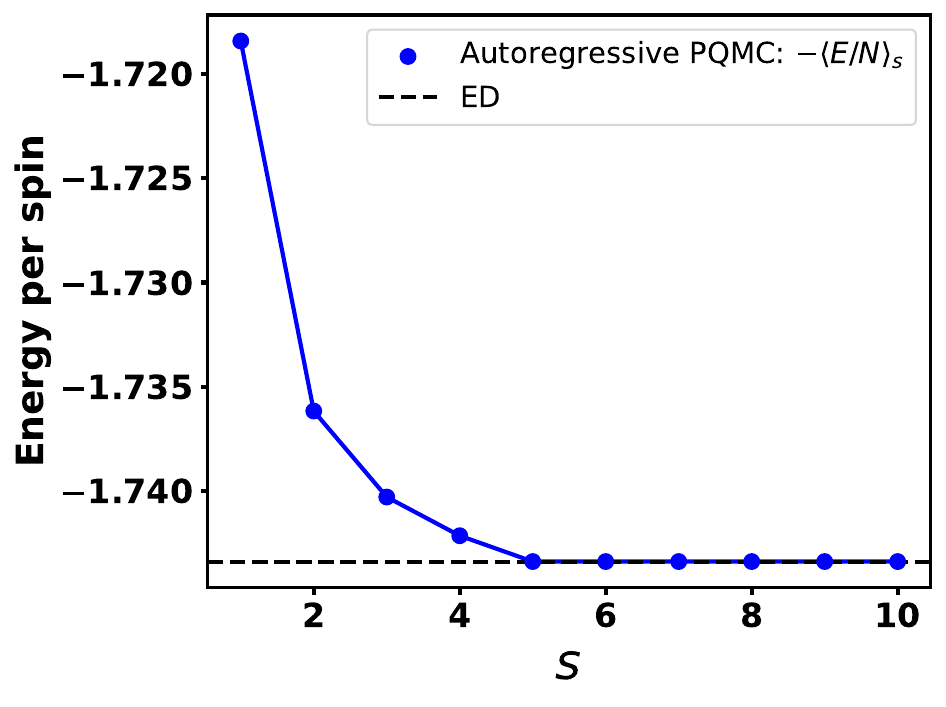} }}
    \caption{\textbf{Energy per stint}. We plot the energy per spin per stint $-\langle E/N\rangle_s$ for the  \textbf{(a)} Hermitian, and \textbf{(b)} non-Hermitian Ising chain. We consider $N=20$ and $N=10$, respectively. We observe that the autoregressive PQMC (in blue) is more accurate after each stint finally converging to the ED (black dashed) result.}
    \label{FA1}
\end{figure*}

Assuming the continuous time approximation and importance sampling, the iterative solution of the Schr\"odinger equation for a Markov process reads
\begin{equation}\label{Eq1.53}
    f\left( {\bm{x},\tau  + \Delta \tau } \right) = \sum\limits_{\bm{x'}} {\tilde G\left( {\bm{x},\bm{x'},\Delta \tau } \right)f\left( {\bm{x'},\tau } \right)}. 
\end{equation}
For a sufficiently small $\delta \tau  = \frac{{\Delta \tau }}{M}$, one can Taylor expand the modified Green's function at linear order as
\begin{align}
\tilde G\left( {\bm{x},\bm{x'},\Delta \tau } \right) &\cong {\left[ {g\left( {\bm{x},\bm{x'},\delta \tau } \right)} \right]^M} \nonumber \\
& \cong {\left[ {{\delta _{\bm{x},\bm{x'}}} - \delta \tau \left( {{H_{\bm{x},\bm{x'}}} - {E_{r}}{\delta _{\bm{x},\bm{x'}}}} \right)\frac{{{\psi _T}\left( \bm{x} \right)}}{{{\psi _T}\left( {\bm{x'}} \right)}}} \right]^M}, \label{Eq1.54}
\end{align}
where
\begin{align} \label{Eq1.55}
  &g\left( {\bm{x},\bm{x'},\delta \tau } \right) = {p_{\bm{x},\bm{x'}}}\sum\limits_x {g\left( {\bm{x},\bm{x'},\delta \tau '} \right)}, \\
& {p_{\bm{x},\bm{x'}}} = \frac{{\left[ {{\delta _{\bm{x},\bm{x'}}} - \delta \tau \left( {{H_{\bm{x},\bm{x'}}} - {E_{ref}}{\delta _{\bm{x},\bm{x'}}}} \right)\frac{{{\psi _T}\left( \bm{x} \right)}}{{{\psi _T}\left( {\bm{x'}} \right)}}} \right]}}{{1 - \delta \tau \left[ {{E_{loc}}\left( {\bm{x'}} \right) - {E_{r}}} \right]}},
\end{align}
with $\tilde G_d\left( {\bm{x},\bm{x'},\Delta \tau } \right) \equiv {p_{\bm{x},\bm{x'}}} $, and $\tilde G_b\left( {\bm{x},\bm{x'},\Delta \tau } \right) \equiv \sum\limits_x {g\left( {\bm{x},\bm{x'},\delta \tau '} \right)}=e ^{ - \delta \tau '\left[ {{E_{loc}}\left( {\bm{x'}} \right) - {E_{r}}} \right]}$.

As mentioned before, the idea behind the continuous time algorithm \cite{becca2017quantum} is to slice the time $\Delta \tau$ into $M$ slices and take $M\ \to \infty $. Then, extract the time $\delta\tau'$ that passes before the transition to the next configuration. This is achieved by sampling from a Poisson distribution, which is suitable provided Poisson processes are stochastic processes that model a sequence of independent events occurring randomly at constant over time, so we set
\begin{equation}\label{Eq3.1}
    \delta \tau ' = \min \left[ {\delta {\tau _t},\frac{{\ln \left( {1 - \xi } \right)}}{{{E_{loc}}\left( {\bm{x'}} \right) - {E_{cl}}\left( {\bm{x'}} \right)}}} \right], 
\end{equation}
with $\xi = \left[ 0 , 1 \right]$ a random uniform number. Note that a similar procedure can be repeated for the NH system.

\section{Energy per stint}\label{A1}
In Fig.~\ref{FA1}, we plot the energy per spin as a function while varying the number of stints $s$, and show that the guiding function becomes progressively more accurate. Convergence is reached when successive stints yield stable RNN parameters and stationary PQMC energy estimates.

\section{The sign problem}\label{A2}

In this appendix, we discuss how  the non-Hermitian Hamiltonian in Eq.~\eqref{Eq28} can be transformed to yield a stoquastic Hamiltonian.

\subsection{Gauge rotation}
In quantum Monte Carlo (VMC, PQMC,...), one samples spin configurations $\bm{x}$ with a positive definite weight $\omega(\bm{x})$. In the computational basis $\sigma^z$, the off-diagonal matrix elements $\langle \bm{x}'|H|\bm{x}\rangle$ of our non-Hermitian Ising chain are complex, leading to complex weights $\omega(\bm{x})$. As a result, the variance of observables grows exponentially with system size, making large-scale simulations impractical---this is the so-called ``sign problem''.

To circumvent this problem, we perform a gauge rotation by applying a local unitary transformation (phase rotation) $U$ on the spin operators, so that all off-diagonal matrix elements become real and non-positive. This yields a stoquastic Hamiltonian in the rotated basis, which is sign-problem free. For each sublatice $j$, we define a rotation angle $\theta_j$ that rotates $\sigma_j^x$ such that $ e^{i\theta_j\sigma^z} \sigma^x e^{-i\theta_j\sigma^z} = \cos(2\theta_j)\,\sigma^x - \sin(2\theta_j)\,\sigma^y$, and $U = \exp(i \theta_j \sigma^z)$ is diagonal  in the $\sigma^z$ basis. All off-diagonal elements pick up a phase factor due to the gauge rotation.

Specifically, for a spin flip at site $j$ (where $\bm{x'}$ differs from $\bm{x}$ by a flip at $j$), the matrix elements transform as
\begin{align}
    \langle\bm{x'}|H_{\text{rot}}|\bm{x}\rangle=\langle\bm{x'}|UHU^{\dagger}|\bm{x}\rangle = e^{i[\alpha(\bm{x'})-\alpha(\bm{x})]}\langle\bm{x'}|H|\bm{x}\rangle,
\end{align}
where the phase difference $\alpha(\bm{x'})-\alpha(\bm{x}) = \pm 2\theta_n$ depends on the flip direction ($+2\theta_j$ if flipping a spin-up to a spin-down, and $-2\theta_j$ otherwise) and $H_{\text{rot}}$ is the Hamiltonian in the rotated gauge. For a term $-g\sigma_j^x$ on sublattice $A$, this means the matrix elements transform as $-g\to-ge^{\pm i2\theta_A}$. Similarly, for a term $-g^*\sigma_j^x$ on sublattice $B$, the matrix elements transform as $-g^* \to-g^*e^{\mp i2\theta_B}$.

Now we choose $\theta_A$ and $\theta_B$ so that all transformed off-diagonal elements are real and non-positive. If we let $g=|g|e^{i\phi}$, and $g^*=|g|e^{-i\phi}$, then for $\theta_A=-\phi/2$ and $\theta_B=\phi/2$ the Hamiltonian is stoquastic and one can verify that both flip directions yield real, non-positive values: Indeed, $-ge^{\pm i2\theta_A}=-|g|e^{i(\phi \mp \phi)}$ and $-g^*e^{\mp i2\theta_B}=-|g|e^{i(-\phi\pm\phi)}$, which equal either $-|g|\leq 0$  or $-|g|e^{\pm i2\phi}$  depending on direction. By the bipartite structure and appropriate choice of phases, the Hamiltonian becomes stoquastic in the later case. Note, however, that the choice of $\pm$ and $\mp$ for sublattices $A$ and $B$, respectively, is just a labeling choice and has nothing to do with the $A/B$ asymmetry. However, this labeling choice is chosen in advance such that once the values of $\theta_A$ and $\theta_B$ are substituted, the upper sign case for $A$ and the upper sign case for $B$ line up to give the matching results described above.

The rotated Hamiltonian reads 
\begin{align}
    H_\textrm{rot}=UHU^{\dagger}; \,\, U=\prod_{j\in A}e^{-i\theta_A\sigma_j^z}\prod_{j\in B}e^{-i\theta_B\sigma_j^z},
\end{align}
with $\theta_A =-\frac{\text{arg}(g)}{2}$ and $\theta_B =\frac{\text{arg}(g)}{2}$, such that $U$ is diagonal in the $\sigma^z$ basis. $U$ leaves the diagonal terms of the Hamiltonian  unchanged.  As this is a unitary transformation, the spectrum of the system is preserved (that is, the spectrum of $H_{\text{rot}}$ and that of $H$ are identical), as well as the $PT$ symmetry even though the $PT$ operator will be modified. The original ground state is recovered via $|\psi\rangle = U^\dagger |\psi_{\mathrm{rot}}\rangle$.

\subsection{Similarity transform}
In addition to the gauge rotation, one could also use a similarity transformation~\cite{mostafazadeh2003exact,MIAO20161805} to mitigate the sign problem in NH systems. In principle, for a NH system in a $PT$-unbroken regime, one can always find a similarity transformation that transforms the model into a Hermitian model~\cite{mostafazadeh2003exact}. 

For a $PT$-symmetric Hamiltonian $H$ in the unbroken phase, there exist a positive-definite metric operator $\zeta$ such that
\begin{align*}
    H^{\dagger}=\zeta H \zeta^{-1}.
\end{align*}
The similarity transformation
\begin{align}\label{B2}
    h=\zeta^{1/2} H \zeta^{-1/2},
\end{align}
yields a Hermitian Hamiltonian 
\begin{align*}
    h=h^{\dagger},
\end{align*}
which is isospectral to $H$ (that is, with the same real eigenvalues as $H$), and the eigenstates are related to each other by $|\psi_n\rangle_H=\zeta^{1/2}|\psi_n\rangle_h$~\cite{mostafazadeh2003exact}.

For our $PT$-symmetric TFIM, the metric operator $\zeta$ has the form
\begin{align}
    \zeta=\exp \left({\sum_{j \in A }\beta\sigma_j^y -\sum_{j \in B }\beta\sigma_j^y}\right)=\prod_{j \in A}e^{\beta\sigma_j^y}\prod_{j \in B}e^{-\beta\sigma_j^y},
\end{align}
where the parameter $\beta$ is determined by the non-Hermitician strength $\xi$. The transformation $e^{\beta\sigma_j^y}$ acts on Pauli operators as
\begin{align*}
    e^{\beta\sigma_j^y} \sigma_j^x e^{-\beta\sigma_j^y} &=\cosh(2\beta)\sigma_j^x +\sinh(2\beta)\sigma_j^z, \\
    e^{\beta\sigma_j^y} \sigma_j^z e^{-\beta\sigma_j^y} &=\cosh(2\beta)\sigma_j^z -\sinh(2\beta)\sigma_j^x.
\end{align*}
The parameter $\beta$ is fixed by requiring $H^{\dagger}=\zeta H\zeta^{-1}$. This condition yields
\begin{align*}
    \tanh(2\beta)=\frac{\xi}{\eta} \implies \beta=\tanh^{-1}\left(\frac{\xi}{\eta}\right)=\frac{1}{4}\ln\left(\frac{\eta+\xi}{\eta-\xi}\right),
\end{align*}
which is true for $|\eta|>|\xi|$. After some calculations and imposing the pseudo-Hermiticity condition, the imaginary terms cancel identically yielding the following Hermitian Hamiltonian in the transformed basis
\begin{align}\label{B3}
    h=-J\sum_{j=1}^N\tilde{\sigma}_j^z \tilde{\sigma}_{j+1}^z -2\sqrt{\eta^2 -\xi^2}\sum_{j=1}^N\tilde{\sigma}_j^x,
\end{align}
where $\tilde{\sigma}_j^{\alpha}=\zeta^{1/2}\sigma_j^{\alpha} \zeta^{-1/2}$ (with $\alpha=x,z$) are the transformed Pauli  operators. The Hamiltonian in Eq.~\eqref{B3} is a Hermitian TFIM with transverse field strength $2\sqrt{\eta^2 -\xi^2}=2|g|$ on all sites, and the Ising interaction picks up corrections from the transformation. The off-diagonal elements of Eq.~\eqref{B3} are real and non-positive, as a result, $h$ is stoquastic.

\subsection{Gauge rotation vs similarity transform: The nonlocality}

While mathematically elegant, we argue that the similarity transform has some practical challenges compared to the gauge rotation method. First, in some cases, the metric operator is constructed from the outer product of the biorthogonal eigenpairs, and necessitates knowledge of all the eigenstates of the Hamiltonian (defeating the purpose of QMC), which becomes intractable for larger systems size. Some less reliable alternative techniques include, but are not limited to, constructing the metric perturbatively $\eta \approx \mathcal{1} + \sum_{k=1}^\infty \eta^{(k)}$ (where each order can be computed perturbatively provided $|g-g^*|<<1$), or iteratively solving $H^{\dagger}=\zeta H \zeta^{-1}$, which is also limited by the system size. In contrast, the gauge rotation only necessitates finding an appropriate basis, where the Hamiltonian is stoquastic.

Second, $\zeta^{1/2}$ is typically a nonlocal operator, which in turns means that $h$ has long-range or complicated interactions that are hard to sample in the QMC. In particular, when one transforms the nearest-neighbor interaction term $\sigma_j^z\sigma_{J+1}^z$ between sublattices A and B, one obtains contributions such as $\zeta^{1/2}\sigma_j^z\sigma_{J+1}^z\zeta^{1/2}$, which is not a simple two-body operator, and the Baker–Campbell–Hausdorff expansion generates multi-body terms, making the effective Hamiltonian nonlocal. In contrast, the rotation gauge approach uses a local operator that only acts on the spin and its nearest-neighbor without creating any entanglement.

Finally, the similarity approach is generally only valid if the system is in the $PT$-unbroken phase. Such constraints do not apply to the gauge rotation approach. While both methods are applicable to our model, we argue that the gauge rotation method is more viable and generalizable because it does not require a Hermitian embedding.

\subsection{Real-imaginary decomposition}
In addition to the two previously proposed methods, one could use the so-called real-imaginary decomposition, where the idea is to perform a mixed estimator and perturbative treatment of $\xi$ rather than an exact non-Hermitian projection. While relatively simpler than the two previous methods, is method is also restricted to only $PT$-symmetric Hamiltonians in the unbroken phase.

Rather than removing the complex phase of $g$ through a gauge rotation, we exploit the additive structure of the Hamiltonian in Eq.~\eqref{Eq28} and split the transverse-field term as
\begin{align}
    -g\sum_{j\in A}\sigma_j^x-g^*\sum_{j\in B}\sigma_j^x=-\eta\sum_{j= 1}^N\sigma_j^x+i\xi\left(\sum_{j\in B}\sigma_j^x-\sum_{j\in A}\sigma_j^x\right),
\end{align}
such that $H=H_{\eta} +i\xi D$, with
\begin{align}
    H_{\eta}=-J \sum_{j=1}^N \sigma_j^z \sigma_{j+1}^z -\eta\sum_{j= 1}^N\sigma_j^x, \,\, D=\sum_{j\in B}\sigma_j^x-\sum_{j\in A}\sigma_j^x.
\end{align}
$H_{\eta}$ is Hermitian, and for $\eta>0$ all off-diagonal elements are $-\eta$ (real and non-positive) in the computational basis with no rotation required. $D$ is purely off-diagonal and enters $H$ only through the imaginary coefficient $i\xi$. $D$ never needs to be made real, because it is not used to define any stochastic process. Indeed, the branching and diffusion Green's function, the single-flip probability $P_f$ and the flip-site selection can be constructed entirely from $H_{\eta}$. In the importance-sample (guided) walk with guiding wavefunction $\psi_T$, the flip ration and total escape rate of configuration $\bm{x}$ remain the same as in Eq.~\eqref{Eq27}, which is real and positive for a positive RNN. So, $H_{\eta}$ alone is sufficient to generate a descent sign-problem free stochastic walk.

In this case, the operator $i\xi D$ never enters the walker dynamics, the escape rate or the branching weights. Its expectation value is instead accumulated as an additional contribution to the local-energy estimator, evaluated on the ensemble of configurations generated by the $H_{\eta}$-only walk. As such the local energy is computed as $E_{\text{loc}}=E_{\text{loc}}^{\text{Re}}+i\xi E_{\text{loc}}^{\text{Im}}$, where $E_{\text{loc}}^{\text{Re}}$ and $E_{\text{loc}}^{\text{Im}}$ are real-valued arrays tracked seperately walker-by-walker. The complex local energy is never explicitly formed and no complex weight ever appears in the simulation. The population control is driven by $E_{\text{loc}}^{\text{Re}}$ relative to the reference energy, and $E_{\text{loc}}^{\text{Im}}$ is reported alongside.

This procedure amounts to sampling the ground state of the manifestly stoquastic Hermitian Hamiltonian $H_{\eta}$ via standard PQMC, and evaluating $\langle i\xi D\rangle$ as a mixed estimator on that ensemble, rather than projecting the full NH Hamiltonian. This is exact when the guiding wavefunction (and the resulting walkers) is a sufficiently good approximation to the ground state of $H$ itself. Further, this approach is also consistent with our restriction to the $PT$-unbroken regime, where the spectrum is real and we treat $\xi$ as a controlled perturbation on top of the $\eta$-driven dynamics.

\section{Hyperparameters}\label{A3}

All the parameters needed to reproduce the results in this work can be found in Table.~\ref{tab1}. All simulations where performed on an HPC cluster using NVIDIA Quadro RTX 6000 GPUs (24\,GB VRAM, CUDA~12.2).

\begin{table*}[!ht]
\centering
\begin{tabular}{|c|c|c|} 
\hline
Methods & Hyperparameters & Entries \\ \hline
&Optimizer &  Adam \\
&RNN cell &  Vanilla/GRU \\
&Seed & 111 \\
&Input dimension & 2 \\
&Sampling & Exact autoregressive \\
&Coupling J  & 1 \\
RNN&System size & [20,150] \\
&Activation function & tanh/Softmax \\
&Learning rate & $10^{-3}$ \\
&Number of hidden units & 32 \\
&Number of layers & 1 \\
&Number of samples & 1024 \\
&Training steps & $2 \times10^3$ \\
&Ansatz & RNN \\
&Offset in the log probability & $10^{-18}$ \\
\hline
&Importance sampling & True\\
Autoregressive PQMC&Time steps $\Delta \tau$ & $[10^{-1}, 10^{-5}]$\\
&Total projection time $\Delta \tau_p$ & 20\\
&Continuous time & True\\
\hline
\end{tabular}
\caption{\textbf{Simulation parameters}. We record all the parameters used to build the RNN and PQMC routines.}
\label{tab1}
\end{table*}

\section{Analytical solution of the Hermitian TFIM}\label{A5}

Here we analytically solve our Hermitian staggered TFIM in Eq.~\eqref{Eq9}.
After rotating our staggered TFIM to the standard TFIM, we will now diagonalize $\widetilde H$; the original ground state is $|\Omega\rangle = U^\dagger R^\dagger |\widetilde\Omega\rangle = U R^\dagger |\widetilde\Omega\rangle$. Now, let us perform a Jordan-Wigner mapping to free fermions
\begin{align*}
    c_j=\Big(\prod_{\ell<j}\sigma_\ell^z\Big)\,\frac{\sigma_j^x-i\sigma_j^y}{2},\qquad
c_j^\dagger=\Big(\prod_{\ell<j}\sigma_\ell^z\Big)\,\frac{\sigma_j^x+i\sigma_j^y}{2},
\end{align*}
for which $\sigma_j^z=1-2\,c_j^\dagger c_j$. A standard calculation yields
    \begin{align*}
    \widetilde H
&= gN \;-\; 2g \sum_j c_j^\dagger c_j \notag \\
&\;-\;\frac{J}{2}\sum_j \Big[(c_j^\dagger c_{j+1}+c_{j+1}^\dagger c_j)
+ (c_j^\dagger c_{j+1}^\dagger+c_{j+1}c_j)\Big],
\end{align*}
where for PBCs one must pick a fermion-parity sector. The ground state lies in the even-parity sector, which imposes anti-periodic momenta
\begin{align*}
    k=\frac{(2m{+}1)\pi}{N},\qquad m=0,1,\dots,\frac{N}{2}-1.
\end{align*}
We Fourier transform $c_j=\frac{1}{\sqrt N}\sum_{k=-n}^n e^{ikj} c_k$ and group $\pm k$ into Nambu spinors $\Psi_k=(c_k,\;c^\dagger_{-k})^T$, such that
\begin{align*}
    \widetilde H&= \sum_{0<k<\pi} \Psi_k^\dagger\,\mathcal{H}_k\,\Psi_k \;+\; \text{const}, \notag\\ 
&\mathcal{H}_k = 2(g-J\cos k)\,\tau_z + 2J\sin k\,\tau_y,
\end{align*}
with $\tau_{y,z}$ Pauli matrices in Nambu space. The Bogoliubov spectrum is
\begin{align*}
    \varepsilon_k =& 2\sqrt{(g - J\cos k)^2 + (J\sin k)^2}\\ \notag
=& 2\sqrt{g^2+J^2-2gJ\cos k}.
\end{align*}
We define the Bogoliubov angle $\theta_k$ by
\begin{align*}
    \tan(2\theta_k)&=\frac{J\sin k}{\,g - J\cos k},\qquad
\cos(2\theta_k)=\frac{g - J\cos k}{\varepsilon_k/2},\notag \\
\sin(2\theta_k)&=\frac{J\sin k}{\varepsilon_k/2},
\end{align*}
and the quasiparticles with new operators
\begin{align*}
    \gamma_k =& \cos\theta_k\, c_k + i\sin\theta_k\, c^\dagger_{-k},\\ \notag
\gamma_{-k} =& \cos\theta_k\, c_{-k} - i\sin\theta_k\, c^\dagger_{k},
\end{align*}
which diagonalize the Hamiltonian $\widetilde H$ as
\begin{equation*}
    \widetilde H = \sum_{0<k<\pi} \varepsilon_k \big(\gamma_k^\dagger\gamma_k + \gamma_{-k}^\dagger\gamma_{-k} - 1\big).
\end{equation*}
Now, we write the exact ground state (Bardeen-Cooper-Schrieffer form) and energy. Let $|0\rangle$ be the fermion vacuum (all spins up along $\sigma^z$ in the rotated frame). The unique even-parity ground state of $\widetilde H$ is the Bogoliubov vacuum $\gamma_{\pm k}|\widetilde\Omega\rangle=0$, that is
\begin{align*}
    |\widetilde\Omega\rangle
= \prod_{0<k<\pi}\Big(\cos\theta_k \;-\; i\sin\theta_k\, c_k^\dagger c_{-k}^\dagger\Big)\,|0\rangle .
\end{align*}
One could take a real gauge $ -i\!\to\! +1$, however, any overall $k$-dependent phase choice is equivalent. The ground-state energy yields
\begin{align}\label{EqA}
    E_0(\widetilde H)&= -\sum_{0<k<\pi}\varepsilon_k
= -\frac{N}{2\pi}\int_0^\pi \varepsilon_k\,dk \notag \\
\quad\Longrightarrow\quad
\frac{E_0}{N} &= -\frac{1}{\pi}\int_0^\pi \sqrt{(g - J\cos k)^2 + (J\sin k)^2}\,dk.
\end{align}

One can check that when $J=0\Rightarrow \varepsilon_k=2g\Rightarrow E_0/N=-g$, that is all spins polarize along $-\hat z$ in the rotated frame (paramagnet). Also, when $g=0\Rightarrow \varepsilon_k=2|J|\Rightarrow E_0/N=-|J|$, the model represents a ferromagnet along $\hat x$ in the rotated frame. Finally, the ground state of our original Hamiltonian is obtained by undoing the rotations
\begin{align*}
    |\Omega\rangle = U^\dagger R^\dagger |\widetilde\Omega\rangle
&= \Big(\prod_{j\ \text{odd}}\sigma_j^z\Big) 
\Big(\prod_{i} e^{+i\frac{\pi}{4}\sigma_i^y}\Big) \notag \\
&\prod_{0<k<\pi}\big(\cos\theta_k - i\sin\theta_k\, c_k^\dagger c_{-k}^\dagger\big)\,|0\rangle. 
\end{align*}
This is an exact, ``closed-form'' expression. It is the usual TFIM Bardeen–Cooper–Schrieffer vacuum, then rotated back and ``de-staggered''. The corresponding exact ground-state energy for our original $H$ is the same as above (since unitaries do not change energies), cf. Eq.~\eqref{EqA}. With PBCs, Jordan–Wigner introduces two fermionic sectors (even/odd parity). The physical ground state is in the even-parity sector, which corresponds to anti-periodic fermions and momenta $k=(2m{+}1)\pi/N$. On a finite ring at the critical point $g=J$, there is a subtle near-degeneracy that disappears as $N\to\infty$; the formulas above remain valid.

For the non-Hermitian TFIM, the exact ground state can be obtained using series expansion method as derived in Ref.\cite{PhysRevB.104.195137}.

\begin{figure}[t]
    \centering
    \includegraphics[width=0.48\textwidth]{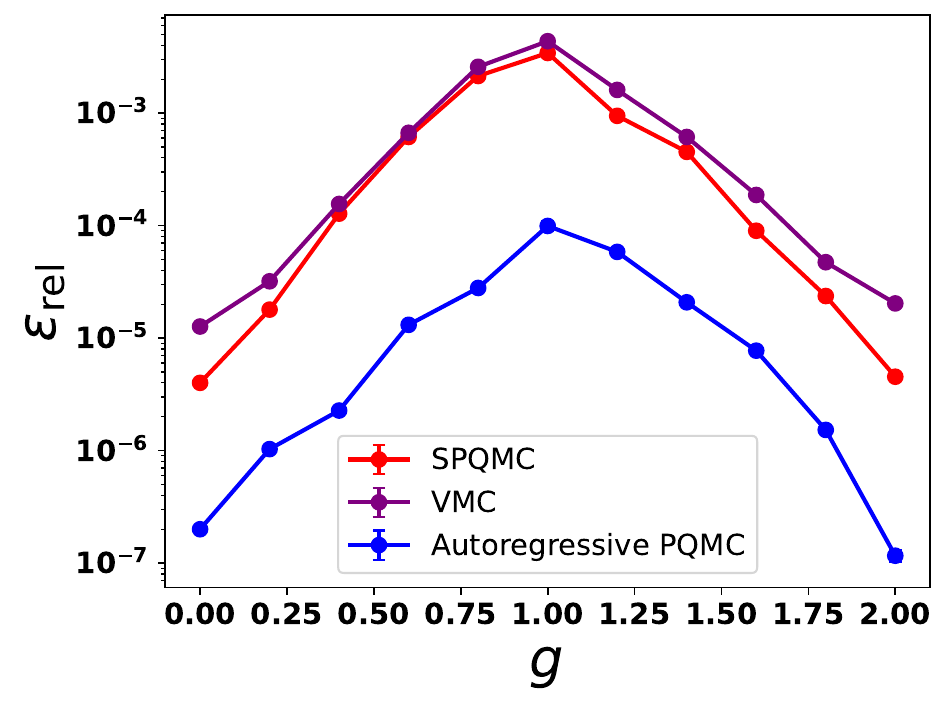}
    \caption{\textbf{Variational RNN versus SPQMC versus autoregressive PQMC}. We compare the variational RNN (RNN with VMC) with the autoregressive PQMC. We see that the later (blue) achieves less error than the variational RNN (purple), which is also less accurate than SPQMC (red). We considered $N=20$ spins for the Hermitian Ising chain, and PBCs. }
    \label{FA2}
\end{figure}

\section{Variational RNN versus projective RNN}\label{A4}

Here we compare the results of RNN with VMC and RNN with PQMC (autoregressive PQMC). We show that autoregressive PQMC is almost two orders of magnitude more accurate than variational RNN. Further, we also note that even SPQMC outperforms the variational RNN cf. Fig.~\ref{FA2}.

\bibliography{bibliography}

\end{document}